\documentclass[aps,prd,reprint,longbibliography,nofootinbib,floatfix,
author-year]{revtex4-2}

\usepackage[T1]{fontenc}
\usepackage{newtxtext,newtxmath}  

\usepackage{graphicx}
\usepackage{float}            

\usepackage{mathtools}
\usepackage{amssymb}
\usepackage{bm}
\usepackage{booktabs}
\usepackage{natbib}
\usepackage[colorlinks=true,linkcolor=blue,citecolor=magenta,urlcolor=blue]{hyperref}

\def\doi#1{\href{https://doi.org/#1}{\color{blue}#1}}

\usepackage{xstring}
\usepackage{tikz}
\usetikzlibrary{shapes.geometric,shapes.symbols}

\makeatletter
\let\ORIGurl\url
\renewcommand{\url}[1]{%
	\begingroup
	\def\UrlArg{#1}%
	\IfBeginWith{\UrlArg}{https://doi.org/}{\href{\UrlArg}{\StrBehind{\UrlArg}{https://doi.org/}}}{%
		\IfBeginWith{\UrlArg}{http://doi.org/}{\href{\UrlArg}{\StrBehind{\UrlArg}{http://doi.org/}}}{%
			\IfBeginWith{\UrlArg}{https://arxiv.org/abs/}{\href{\UrlArg}{arXiv:\StrBehind{\UrlArg}{https://arxiv.org/abs/}}}{%
				\IfBeginWith{\UrlArg}{http://arxiv.org/abs/}{\href{\UrlArg}{arXiv:\StrBehind{\UrlArg}{http://arxiv.org/abs/}}}{%
					\IfBeginWith{\UrlArg}{https://ui.adsabs.harvard.edu/abs/}{\href{\UrlArg}{\StrBehind{\UrlArg}{https://ui.adsabs.harvard.edu/abs/}}}{%
						\IfBeginWith{\UrlArg}{http://ui.adsabs.harvard.edu/abs/}{\href{\UrlArg}{\StrBehind{\UrlArg}{http://ui.adsabs.harvard.edu/abs/}}}{%
							\ORIGurl{\UrlArg}%
	}}}}}}%
	\endgroup
}
\makeatother

\definecolor{mkOrange}{HTML}{E69F00}
\definecolor{mkBlue}{HTML}{0173B2}
\definecolor{mkRed}{HTML}{D55E00}
\definecolor{mkGreen}{HTML}{029E73}
\definecolor{mkPurple}{HTML}{CC78BC}

\newcommand{\mpsi}{m_\psi}
\newcommand{\mchi}{m_\chi}
\newcommand{\mphi}{m_\phi}
\newcommand{\gphi}{g_\phi}
\newcommand{\gchi}{g_\chi}
\newcommand{\lamB}{\lambda_{\mathrm{B}}}
\newcommand{\eV}{\mathrm{eV}}
\newcommand{\Msun}{M_\odot}

\newcommand{\kappaphi}{\kappa_\phi}
\newcommand{\RT}{R_T}
\newcommand{\xT}{x_T}

\DeclareRobustCommand{\orcidicon}{%
	\begin{tikzpicture}
		\definecolor{orcidgreen}{HTML}{166B3A}
		\draw[orcidgreen, fill=orcidgreen] (0,0)
		circle [radius=0.16]
		node[white] {{\fontfamily{qag}\selectfont \tiny ID}};
	\end{tikzpicture}
	\hspace{-2mm}
}
\foreach \x in {A, ..., Z}{%
	\expandafter\xdef\csname orcid\x\endcsname{\noexpand\href{https://orcid.org/\csname orcidauthor\x\endcsname}{\noexpand\orcidicon}}
}

\begin{document}
	\flushbottom

	\title{General-relativistic structure of \\ two-component quantum dark fermion stars\thanks{Published version: I.~Lopes, Phys.\ Rev.\ D \textbf{114}, 035032 (2026), \doi{10.1103/xzfb-4mrn}.}}

	\author{Il\'idio Lopes\,\orcidA{}}
	\affiliation{\href{https://ror.org/020hp3377}{Centro de Astrof\'isica e
Gravita\c{c}\~ao (CENTRA)}, Departamento de F\'isica, \\
Instituto Superior T\'ecnico (IST),
\href{https://ror.org/01c27hj86}{Universidade de Lisboa (UL)}, \\
Av.~Rovisco Pais~1, 1049-001 Lisboa, Portugal}

	\received{24 May 2026}
	\accepted{23 July 2026}
	\published{25 August 2026}

\begin{abstract}
We develop a general-relativistic framework for two-component quantum dark fermion stars: equilibrium configurations of two degenerate fermion species governed jointly by gravity, a Yukawa-mediated dark fifth force, and a globally relevant Bohm quantum-pressure correction.
The treatment retains the full covariant form of the nonlinear Klein--Gordon equation in the Schwarzschild interior, with closure relations valid at arbitrary compactness, from the ultralight baseline up to densities of order $0.16\,\mathrm{fm}^{-3}$ at which relativistic scalar densities and self-consistent effective fermion masses become unavoidable.
Two Lagrangian parameters, the dark-fermion mass and the dimensionless ratio between the Yukawa channel and gravity, together fix the equilibrium structure; a joint measurement of mass and radius therefore constrains the dark-sector microphysics directly from gravitational-wave observables.
For particle masses in the band $10^{-11}$--$10^{-10}\,\eV$, the configurations exhibit radii of $3$--$24\,\mathrm{km}$ and compactness in the range $0.14$--$0.34$, behaving as dual mimickers: at moderate compactness they overlap with neutron stars in mass, radius, and inspiral frequency; at the most compact end ($\kappa\sim 0.34$, $R_T/R_S\simeq 1.5$) they cross the photon sphere and could masquerade as low-mass black holes in the mass-gap region.
Tidal-deformability measurements discriminate against both populations simultaneously: the dimensionless tidal deformability is measurably smaller than the neutron-star value yet remains nonzero, unlike that of a genuine black hole.
These objects populate the sensitivity bands of LISA, the Einstein Telescope, and Cosmic Explorer, producing astrometric microlensing signatures individually resolvable by \textit{Gaia}.
The framework thus provides a reproducible and falsifiable template for constraining dark-sector properties through multimessenger observations.
\end{abstract}

	\maketitle

	\section{Introduction}
	\label{sec:intro}

	The observational evidence for nonbaryonic dark matter, assembled from independent measurements of the cosmic microwave background, galaxy rotation curves, and strong-lensing observations, establishes that roughly $27\%$ of the cosmic energy density resides in a gravitationally active yet electromagnetically inert component whose particle-physics identity remains unknown \citep{2024PhRvD.110c0001N,2020A&A...641A...6P}.
	In the ultralight regime ($m\lesssim 1\,\eV$), quantum effects manifest on astrophysical scales, giving rise to solitonic structures whose characteristic size is set by the de~Broglie wavelength \citep{2014NatPh..10..496S,2017PhRvD..95d3541H,2021A&ARv..29....7F}, while phase-space constraints on fermionic dark matter in dwarf galaxies place lower bounds on the particle mass that depend sensitively on the number of spin states and the formation history \citep{1979PhRvL..42..407T}.
	Were a fraction of this component to reside in stable, self-gravitating compact objects rather than being distributed diffusely, the dark matter problem would partly become a matter of compact-object astrophysics, with observational signatures accessible to gravitational-wave interferometers and microlensing surveys.

	The equilibrium structure of two-component dark-fermion configurations has been studied within the Newtonian quantum-hydrodynamic (Madelung) framework, where the Bohm potential contributes a species-dependent surface-energy correction in the Thomas--Fermi regime \citep{1987PhRvD..35.3678L,2023PhRvD.108d4024D}.
	For dark-fermion masses $m_1\gtrsim 10^{-11}\,\eV$ and total masses $M\sim 0.1$--$5\,\Msun$, the characteristic radii descend to $\sim 10^{2}\,\mathrm{km}$, producing compactnesses $\kappa\sim 10^{-3}$--$10^{-1}$ and gravitational-wave contact frequencies squarely within the sensitivity bands of next-generation detectors \citep{2017arXiv170200786A,2010CQGra..27s4002P,2019BAAS...51g..35R}.
	At such compactnesses, the full Tolman--Oppenheimer--Volkoff structure becomes the natural starting point rather than a perturbative correction, and the question of interest shifts from quantifying departures from Newtonian gravity to determining what novel equilibrium configurations the complete framework predicts.

	The theoretical framework for self-gravitating quantum systems in general relativity has deep roots.
	\citet{1939PhRv...55..374O} established the mass limit of cold degenerate neutron stars, \citet{1968PhRv..172.1331K} studied the Klein--Gordon geon, \citet{1969PhRv..187.1767R} formulated the general-relativistic theory of boson stars, and \citet{1986PhRvL..57.2485C} demonstrated that repulsive self-interactions can raise the maximum boson-star mass to astrophysically interesting values \citep[see][for a comprehensive review]{2023LRR....26....1L}.
	More recently, the TOV formalism has been extended to two-fluid systems in which dark matter and ordinary matter share a common metric while obeying independent equations of state \citep{2024PhR..1052....1B,2024PhLB..85638901P,2025PDU....49Panotopoulos,2020PhRvD.102f3028I}, the structural properties of dark-matter-admixed strange quark stars have been explored \citep{2017PhRvD..96h3013P,2018PhRvD..97b4030L}, and the tidal-deformability consequences for gravitational-wave signals have been investigated \citep{2010PhRvD..81l3521D,2011PhRvD..83h3512K,2026PhRvD.113d3049B}.

	Throughout this paper, we shall refer to such configurations as \emph{two-component quantum dark fermion stars}, the qualifier ``quantum'' denoting the globally relevant Bohm gradient term \citep{2026PhRvD.113k5043L} that enters the equilibrium equations on the same footing as gravity and degeneracy pressure, rather than being confined to a thin surface layer or neglected altogether, as in earlier dark fermion star treatments \citep{1987PhRvD..35.3678L,2023PhRvD.108d4024D}.\footnote{The qualifier ``quantum'' here is intended to distinguish these configurations from classical fermion star treatments in which Bohm gradient corrections are neglected, and not to suggest kinship with boson stars, $Q$-balls, or fermion soliton stars, which differ both in field content and in equilibrium mechanism.}
	The present work addresses the strong-field regime, where the covariant Laplacian governs the Bohm contribution and the Tolman--Oppenheimer--Volkoff structure replaces the Newtonian hydrostatic balance.

	In this paper, I develop a fully general-relativistic framework for quantum dark fermion stars in which gravity and a Yukawa-mediated dark fifth force jointly govern equilibrium, with explicit treatment of internal densities comparable to those encountered in neutron stars ($n\sim 0.16\,\mathrm{fm}^{-3}$, $\kappa\sim 0.1$--$0.3$).
	The principal original contributions are:
	(i)~the exact static Klein--Gordon equation in the Schwarzschild interior, retaining the full covariant Laplacian and nonlinear scalar self-interactions, with Lorentz-scalar sources computed as relativistic Walecka scalar densities at nuclear densities;
	(ii)~the coupling ratio $\beta_i\equiv g_i^2/(4\pi Gm_i^2)$ and two-body interaction potential, which quantify the interplay between gravitational and Yukawa channels;
	(iii)~modified TOV equations incorporating the Bohm quantum-pressure correction evaluated with the covariant Laplacian in the interior Schwarzschild metric;
	(iv)~a numerical algorithm for the general dimensionless boundary-value problem, validated against established Newtonian benchmarks and extended to high-compactness configurations ($\kappa\sim 0.1$--$0.3$);
	(v)~the complete boundary-value problem for nuclear-density quantum dark fermion stars, including a strategy for constraining the self-coupling parameters and a treatment of the ultrarelativistic $\gamma=4/3$ limit.

	The remainder of this paper is organized as follows.
	Section~\ref{sec:model} derives the exact static Klein--Gordon equation in curved spacetime and introduces the coupling ratio~$\beta_i$.
	Sections~\ref{sec:GR}--\ref{sec:numerics} assemble the equilibrium structure equations, the dimensionless formulation, and the numerical implementation.
	Section~\ref{sec:results} presents the ultralight baseline together with the nuclear-density extension; Section~\ref{sec:observations} translates the equilibrium solutions into gravitational-wave, mass--radius, and microlensing predictions; and Section~\ref{sec:conclusions} draws together the conclusions.

	We adopt natural units ($\hbar=c=1$) where convenient, restoring explicit factors as needed for physical clarity.
	Standard conversion factors and the IAU~2015 nominal solar values are used throughout \citep{2024PhRvD.110c0001N,2016AJ....152...41P}.

	\section{Spacetime geometry and matter Lagrangian}
	\label{sec:model}

	In this section I state the Einstein field equations and the interior metric appropriate to a static, spherically symmetric quantum dark fermion star, then introduce the dark-sector field content whose stress-energy tensor sources the gravitational field.
	The Lagrangian is written in its generally covariant form and the field equations are derived; the section concludes with the nonrelativistic reduction, the approximation chain, and the two-body interaction potential.

	\subsection{Einstein equations and interior metric}
	\label{sec:einstein_metric}

	The gravitational dynamics are governed by the Einstein field equations
	\begin{equation}
		G_{\mu\nu} = \frac{8\pi G}{c^4}\,T_{\mu\nu}\,,
		\label{eq:Einstein}
	\end{equation}
	where $G_{\mu\nu}=R_{\mu\nu}-\tfrac{1}{2}Rg_{\mu\nu}$ is the Einstein tensor and $T_{\mu\nu}$ is the total stress-energy tensor of the matter content.
	Here $R_{\mu\nu}=R^{\alpha}{}_{\mu\alpha\nu}$ is the Ricci curvature tensor, obtained by contracting the Riemann tensor of the metric $g_{\mu\nu}$, and $R=g^{\mu\nu}R_{\mu\nu}$ is the Ricci scalar, with the sign conventions of \citet{1973grav.book.....M}; $G$ and $c$ denote Newton's gravitational constant and the speed of light, respectively. A glossary of all symbols and rescaled quantities used below is given in Appendix~\ref{app:notation}.
	The system comprises two dark-sector spin-$\tfrac{1}{2}$ fermion species, $\psi$ (mass $m_1\equiv\mpsi$) and $\chi$ (mass $m_2\equiv\mchi$), coupled through a real scalar mediator $\phi$ (mass $\mphi$); the total stress-energy tensor accordingly decomposes as
	\begin{equation}
		T_{\mu\nu} = T_{\mu\nu}^{(\psi)} + T_{\mu\nu}^{(\chi)} + T_{\mu\nu}^{(\phi)}\,,
		\label{eq:Tmunu_decomp}
	\end{equation}
	where each contribution is determined by the matter Lagrangian specified below.
	It is worth noting that the two species interact with one another only through the shared mediator $\phi$ and through the gravitational field itself; no direct four-fermion coupling is assumed.

	For a static, spherically symmetric configuration the line element takes the Schwarzschild-like form
	\begin{equation}
		\mathrm{d}s^2 = -e^{2\nu(r)}\,c^2\,\mathrm{d}t^2 + \left(1-\frac{2Gm(r)}{rc^2}\right)^{\!-1}\mathrm{d}r^2 + r^2\,\mathrm{d}\Omega^2\,,
		\label{eq:metric}
	\end{equation}
	where $\nu(r)$ is the metric potential, $m(r)$ the enclosed gravitational mass, and $\mathrm{d}\Omega^2=\mathrm{d}\theta^2+\sin^2\!\theta\,\mathrm{d}\varphi^2$.
	Throughout this work we adopt the mostly-plus metric signature $(-,+,+,+)$ \citep{1973grav.book.....M}; in particular, the scalar-field Lagrangian carries the sign $-\tfrac{1}{2}g^{\mu\nu}\partial_\mu\phi\,\partial_\nu\phi$ so that the kinetic energy is positive-definite and the static Klein--Gordon equation admits the Yukawa solution $\phi\propto e^{-\mphi r}/r$.
	It is convenient to define the metric function
	\begin{equation}
		A(r) \equiv e^{-2\lambda(r)} \equiv 1 - \frac{2Gm(r)}{rc^2}\,,
		\label{eq:metric_function}
	\end{equation}
	so that $g_{rr}=A^{-1}=e^{2\lambda}$.
	Matching to the exterior Schwarzschild solution at the stellar surface $r=R_T$ determines $\nu(R_T)=\tfrac{1}{2}\ln(1-2GM/R_T c^2)$, where $M=m(R_T)$ is the total gravitational mass.

	With this metric ansatz, the Einstein equations~\eqref{eq:Einstein} reduce to three independent component equations.
	The $(t,t)$ component yields the mass equation, the $(r,r)$ component yields the equation for the metric potential gradient $\mathrm{d}\nu/\mathrm{d}r$, and covariant conservation $\nabla_\mu T^{\mu\nu}=0$ yields the TOV pressure-balance equation; these are derived explicitly once the equation of state has been specified.

	\subsection{Yukawa Lagrangian in curved spacetime}
	\label{sec:yukawa_lagrangian}

	The two dark-sector species $\psi$ and $\chi$ introduced above are coupled to the common scalar mediator $\phi$ through Yukawa vertices of strengths $\gphi$ and $\gchi$.
	In the spacetime~\eqref{eq:metric}, the action reads $S=\int\!\mathrm{d}^4x\,\sqrt{-g}\,\mathcal{L}$, where the Lagrangian density, standard in nuclear mean-field models of the Walecka type \citep{1997IJMPE...6..515S} and written here in its generally covariant form following the curved-spacetime prescriptions of \citet{1996qtf..book.....B,2009qftc.book.....P}, is
	\begin{equation}
		\begin{aligned}
			\mathcal{L}={}&-\frac{1}{2}\,g^{\mu\nu}\partial_\mu\phi\,\partial_\nu\phi - V(\phi) +\bar\psi\,(i\gamma^\mu D_\mu-m_1)\psi\\
			&+\bar\chi\,(i\gamma^\mu D_\mu-m_2)\chi -\gphi\,\bar\psi\psi\,\phi -\gchi\,\bar\chi\chi\,\phi\,,
		\end{aligned}
		\label{eq:Lagrangian}
	\end{equation}
	with most general renormalizable scalar potential
	\begin{equation}
		V(\phi)=\frac{1}{2}\mphi^2\phi^2
		+\frac{\kappa_3}{3!}\phi^3
		+\frac{\lambda_\phi}{4!}\phi^4\,,
		\qquad \lambda_\phi>0\,.
		\label{eq:Vphi}
	\end{equation}
	Here $D_\mu=\partial_\mu+\Omega_\mu$ is the spinor covariant derivative and $\gamma^\mu=e^\mu_{\ a}\gamma^a$ are the curved-spacetime Dirac matrices.
	The vierbein (tetrad) $e^{\ a}_{\mu}$ and its inverse $e^\mu_{\ a}$ relate the spacetime metric to the local Minkowski metric $\eta_{ab}=\mathrm{diag}(-1,+1,+1,+1)$ through $g_{\mu\nu}=\eta_{ab}\,e^{\ a}_{\mu}e^{\ b}_{\nu}$; the $\gamma^a$ are the constant flat-space Dirac matrices with $\{\gamma^a,\gamma^b\}=2\eta^{ab}$. The spin (Fock--Ivanenko) connection $\Omega_\mu=\tfrac{1}{4}\,\omega_{\mu}{}^{ab}\,\gamma_a\gamma_b$ \citep{1929ZPhy...56..330W,1929ZPhy...57..261F}, built from the Ricci rotation coefficients $\omega_{\mu}{}^{ab}$, renders $D_\mu\psi$ covariant under local Lorentz rotations of the tetrad frame.
	The Yukawa couplings explicitly break $\phi\to-\phi$, permitting the cubic interaction; boundedness from below requires $\lambda_\phi>0$.

	Variation of the action yields the covariant Klein--Gordon equation
	\begin{equation}
		\Box_g\,\phi - \mphi^2\phi
		-\frac{\kappa_3}{2}\phi^2-\frac{\lambda_\phi}{6}\phi^3
		=\gphi\,\bar\psi\psi + \gchi\,\bar\chi\chi\,,
		\label{eq:EL_KG}
	\end{equation}
	where $\Box_g\,\phi$ is the covariant d'Alembertian, and the curved-spacetime Dirac equations in the scalar background are
	\begin{align}
		(i\gamma^\mu D_\mu - m_1)\psi &= \gphi\,\phi\,\psi\,,\label{eq:Dirac_psi}\\
		(i\gamma^\mu D_\mu - m_2)\chi &= \gchi\,\phi\,\chi\,.
		\label{eq:Dirac_chi}
	\end{align}
	The right-hand sides may be absorbed into effective position-dependent masses, $m_1^*(r) = m_1 + \gphi\,\phi(r)$ and $m_2^*(r) = m_2 + \gchi\,\phi(r)$, so the scalar mediator dresses each fermion species with a spatially varying mass shift.
	For the static metric~\eqref{eq:metric}, the d'Alembertian reduces to the covariant Laplacian.  Writing the radial part of $\Box_g\phi$ explicitly,
	\begin{align}
		\nabla^2_{\mathrm{cov}}\phi
		&= \frac{1}{\sqrt{-g}}\,\partial_r\!\bigl(\sqrt{-g}\,g^{rr}\,\partial_r\phi\bigr) \nonumber\\
		&= A\!\left[\phi'' + \left(\frac{2}{r}+\nu'-\lambda'\right)\phi'\right],
		\label{eq:box_to_cov}
	\end{align}
	where $\sqrt{-g}=e^{\nu+\lambda}r^2\sin\theta$, $g^{rr}=A=e^{-2\lambda}$, and $A(r)$ is given by Eq.~\eqref{eq:metric_function}.
	The additional coefficient $(\nu'-\lambda')$ acts as a position-dependent friction that modifies the scalar-field propagation, much as a spatially varying refractive index alters the propagation of light through an inhomogeneous medium.
	This gives the exact static Klein--Gordon equation in the Schwarzschild interior,
	\begin{equation}
		\nabla^2_{\mathrm{cov}}\phi - \mphi^2\phi
		-\frac{\kappa_3}{2}\phi^2-\frac{\lambda_\phi}{6}\phi^3
		=\gphi\,\bar\psi\psi + \gchi\,\bar\chi\chi\,.
		\label{eq:KG_curved}
	\end{equation}
	It is worth emphasising that Eq.~\eqref{eq:KG_curved} is exact at arbitrary compactness: the covariant Laplacian encodes the full interior metric, no weak-field reduction is imposed, and the cubic and quartic self-interactions are retained in their entirety.

	At nuclear densities, where the Fermi momentum becomes comparable to the particle mass, the Lorentz-scalar source $\langle\bar\psi_i\psi_i\rangle$ must be evaluated relativistically.
	In the Walecka mean-field approximation, each species acquires an effective mass
	\begin{equation}
		m_i^*(r) = m_i + g_i\,\phi(r)\,,\qquad g_1\equiv\gphi,\;g_2\equiv\gchi\,,
		\label{eq:mstar_def}
	\end{equation}
	and the scalar density is given by the momentum integral
	\begin{equation}
		n_{s,i}(r) = \frac{1}{\pi^2}\int_0^{p_{F,i}}\!\frac{m_i^*(r)}{\sqrt{p^2+m_i^{*2}(r)}}\;p^2\,\mathrm{d}p\,,
		\label{eq:ns_walecka}
	\end{equation}
	with Fermi momentum related to the number density by $n_i = p_{F,i}^3/(3\pi^2)$.
	The total relativistic energy density and pressure, summed over all species, read
	\begin{align}
		\varepsilon_{\mathrm{tot}} &= \sum_i\frac{1}{\pi^2}\int_0^{p_{F,i}}\!\sqrt{p^2+m_i^{*2}}\;p^2\,\mathrm{d}p + V(\phi)\,,
		\label{eq:eps_RMF}\\
		P_{\mathrm{tot}} &= \sum_i\frac{1}{3\pi^2}\int_0^{p_{F,i}}\!\frac{p^4}{\sqrt{p^2+m_i^{*2}}}\,\mathrm{d}p - V(\phi)\,,
		\label{eq:P_RMF}
	\end{align}
	where the scalar-potential contributions $\pm V(\phi)$ enter the totals once, not per species.
	Replacing the operator bilinears in Eq.~\eqref{eq:KG_curved} by $n_{s,i}$ yields the central field equation of this work, the nonlinear Klein--Gordon equation with fully relativistic source terms:
	\begin{equation}
		(\nabla^2_{\mathrm{cov}}-\mphi^2)\phi
		-\frac{\kappa_3}{2}\phi^2 - \frac{\lambda_\phi}{6}\phi^3
		=\gphi\,n_{s,1} + \gchi\,n_{s,2}\,.
		\label{eq:KG_nuclear}
	\end{equation}
	The self-consistency loop is closed by Eq.~\eqref{eq:mstar_def}: at each radial point the effective mass depends on the local field amplitude, which in turn depends on the scalar densities that are themselves functionals of $m_i^*$.
	At saturation density ($n\simeq 0.16\,\mathrm{fm}^{-3}$), the ratio $n_{s,i}/n_i\simeq m_i^*/E_{F,i}^* \sim 0.6$--$0.7$, so the scalar source is appreciably weaker than the number density.

	Before proceeding to the equilibrium structure equations, it is useful to delineate the approximation chain solved numerically in Section~\ref{sec:numerics}; three successive simplifications may be distinguished.
	The most general case retains relativistic scalar densities $n_{s,i}$~\eqref{eq:ns_walecka}, which couple to the Klein--Gordon equation through Eq.~\eqref{eq:KG_nuclear}.
	In the nonrelativistic limit, the scalar density reduces to the number density, $n_{s,i}\to n_i$, and Eq.~\eqref{eq:KG_nuclear} becomes
	\begin{equation}
		(\nabla^2_{\mathrm{cov}}-\mphi^2)\phi
		-\frac{\kappa_3}{2}\phi^2 - \frac{\lambda_\phi}{6}\phi^3
		=\gphi\,n_1 + \gchi\,n_2\,.
		\label{eq:KG_full_main}
	\end{equation}
	For ultralight dark-sector configurations, the self-coupling terms are negligible and Eq.~\eqref{eq:KG_full_main} reduces further to the linear mediator equation
	\begin{equation}
		(\nabla^2_{\mathrm{cov}}-\mphi^2)\phi
		=\gphi\,n_1 + \gchi\,n_2\,,
		\label{eq:KG_linear}
	\end{equation}
	which constitutes the working equation for the ultralight baseline explored in Section~\ref{sec:results}.

	The effective interaction energy between species $i$ and $j$ separates into gravitational and Yukawa channels:
	\begin{equation}
		\mathcal{V}_{ij}(r) = m_j\,c^2\bigl(e^{\nu(r)}-1\bigr) - \frac{g_ig_j}{4\pi}\,\frac{e^{-\mphi r}}{r}\,,
		\label{eq:Vij}
	\end{equation}
	with $g_1=\gphi$ and $g_2=\gchi$ as before.
	The relative importance of the two channels for species $i$ is measured by the dimensionless coupling ratio
	\begin{equation}
		\beta_i \equiv \frac{g_i^2}{4\pi G m_i^2}\,,
		\label{eq:beta_def}
	\end{equation}
	where $\beta_i=1$ corresponds to a Yukawa coupling of comparable strength to gravity, $\beta_i\gg 1$ indicates a strongly supergravitational dark fifth force, and the purely gravitational limit is recovered as $\beta_i\to 0$.
	It is worth noting that the cross-interaction between species~1 and~2 ($\propto g_1 g_2$) need not be introduced as a separate term: because both species couple to the same mediator~$\phi$ through the Lagrangian~\eqref{eq:Lagrangian}, the shared scalar field automatically generates the $1$--$1$, $2$--$2$, and $1$--$2$ pairwise channels upon integration of the field equations.

	\section{Equilibrium structure equations}
	\label{sec:GR}

	\subsection{Two-fluid TOV equations and Bohm potential}

	For the metric~\eqref{eq:metric} and the two-species stress-energy tensor, the Einstein field equations yield the mass equation
	\begin{equation}
		\frac{\mathrm{d}m}{\mathrm{d}r} = \frac{4\pi r^2\,\varepsilon_{\mathrm{tot}}}{c^2}\,,
		\label{eq:mass_eq}
	\end{equation}
	where $\varepsilon_{\mathrm{tot}}=\varepsilon_1+\varepsilon_2$, and the TOV equation for the total pressure $P_{\mathrm{tot}}=P_1+P_2$,
	\begin{equation}
		\frac{\mathrm{d}P_{\mathrm{tot}}}{\mathrm{d}r} = -\frac{G}{c^2}\,\frac{(\varepsilon_{\mathrm{tot}}+P_{\mathrm{tot}})}{r^2}\,\frac{\bigl(m+4\pi r^3 P_{\mathrm{tot}}/c^2\bigr)}{1-2Gm/(rc^2)}\,.
		\label{eq:TOV}
	\end{equation}
	For the two-fluid system coupled only gravitationally, each species satisfies its own pressure gradient,
	\begin{equation}
		\frac{\mathrm{d}P_i}{\mathrm{d}r} = -(\varepsilon_i+P_i)\,\frac{\mathrm{d}\nu}{\mathrm{d}r}\,,\qquad i\in\{1,2\}\,,
		\label{eq:TOV_species}
	\end{equation}
	which follows exactly from covariant conservation of each species' stress-energy tensor under the shared metric.

	When the dispersive quantum stress is included, the Bohm force supplements the classical pressure gradient and gravitational terms. A self-contained derivation from covariant conservation of the stress-energy tensor is given in Appendix~\ref{app:bohm_derivation}; the result is
	\begin{equation}
		\frac{\mathrm{d}P_i}{\mathrm{d}r} = -(\varepsilon_i+P_i)\,\frac{\mathrm{d}\nu}{\mathrm{d}r} - n_i\,\frac{\mathrm{d}Q_i^{\mathrm{GR}}}{\mathrm{d}r}\,,\qquad i\in\{1,2\}\,,
		\label{eq:TOV_bohm}
	\end{equation}
	where the curved-spacetime Bohm potential is
	\begin{equation}
		Q_i^{\mathrm{GR}} = -\frac{\lamB\hbar^2}{2m_i}\,\frac{\nabla^2_{\mathrm{cov}}\!\sqrt{n_i}}{\sqrt{n_i}}\,,\qquad i\in\{1,2\}\,,
		\label{eq:bohm_GR}
	\end{equation}
	with $\lamB=1/9$ from the Kirzhnits gradient expansion \citep{1935ZPhy...96..431W,2018PhPl...25c2115M,2025PhyU...68..691P}.
	For the metric~\eqref{eq:metric}, the covariant Laplacian of a radial scalar function $f(r)$ is
	\begin{equation}
		\nabla^2_{\mathrm{cov}} f = A\!\left[\frac{\mathrm{d}^2 f}{\mathrm{d}r^2} + \left(\frac{2}{r}+\frac{\mathrm{d}\nu}{\mathrm{d}r}-\frac{\mathrm{d}\lambda}{\mathrm{d}r}\right)\frac{\mathrm{d}f}{\mathrm{d}r}\right].
		\label{eq:cov_laplacian}
	\end{equation}
	The additional coefficient $({\mathrm{d}\nu}/{\mathrm{d}r}-{\mathrm{d}\lambda}/{\mathrm{d}r})$ in Eq.~\eqref{eq:cov_laplacian} encodes the gravitational redshift ($\nu'$) and the radial stretching of proper distance ($\lambda'$); together they modify the kinetic (surface-energy) contribution to the Bohm potential by an amount that is formally $\mathcal{O}(\alpha)$ but becomes significant at high compactness.

	\subsection{Closure relations}

	For each species, I adopt a polytropic equation of state of the form $P_i = K_{p,i}\,n_i^{\gamma_{p,i}}$ \citep[see e.g.][]{2012sse..book.....K}, with $\gamma_{p,i}=5/3$ for a nonrelativistic degenerate Fermi gas and $\gamma_{p,i}=4/3$ for the ultrarelativistic limit.
	One defines the pseudoenthalpy (pressure integral per particle) as $U_{P,i}(n_i)\equiv\int_0^{n_i}\!(\mathrm{d}P_i/\mathrm{d}n_i')\,\mathrm{d}n_i'/n_i'$, which for the polytropic law gives $U_{P,i}=K_{p,i}\gamma_{p,i}\,n_i^{\gamma_{p,i}-1}/(\gamma_{p,i}-1)$.
	The closure relation is obtained by radially integrating the modified TOV equation~\eqref{eq:TOV_bohm}.
	In that equation, the pressure gradient $\mathrm{d}P_i/\mathrm{d}r$ can be rewritten as $n_i\,\mathrm{d}U_{P,i}/\mathrm{d}r$ via the thermodynamic identity $\mathrm{d}P_i=n_i\,\mathrm{d}U_{P,i}$; dividing through by $n_i$ and integrating from the center to radius $r$, one collects the three contributions, the pseudoenthalpy $U_{P,i}$, the gravitational redshift term $m_i c^2(e^{\nu(r)}-1)$, and the Bohm potential $Q_i^{\mathrm{GR}}$, into the algebraic first integral
	\begin{equation}
		\mu_i\,e^{\nu(r)} = U_{P,i}(n_i) + m_i c^2\bigl(e^{\nu(r)}-1\bigr) + Q_i^{\mathrm{GR}}\,,
		\label{eq:GR_closure}
	\end{equation}
	where $\mu_i$ is the integration constant, identified as the Tolman-redshifted chemical potential, which remains constant throughout the star at thermodynamic equilibrium.
	The left-hand side encodes the relativistic condition that the locally measured chemical potential, blueshifted by $e^{\nu}$, is uniform throughout the configuration; the first term on the right is the kinetic (degeneracy) contribution, the second captures the gravitational redshift of the rest-mass energy, and the third is the dispersive Bohm correction.
	In the Newtonian limit ($e^{\nu}\to 1+\Phi/c^2$, $Q_i^{\mathrm{GR}}\to Q_i$), Eq.~\eqref{eq:GR_closure} reduces to the familiar quantum-hydrodynamic Bernoulli equation.

	For the ultralight fermions of interest here ($m_1\sim 10^{-11}\,\eV$), the standard Chandrasekhar bound for the $\gamma_{p,i}=4/3$ polytrope \citep{1939PhRv...55..374O,1983bhwd.book.....S} scales as $m_i^{-2}$ and accordingly lies many orders of magnitude above any astrophysically relevant configuration; it therefore imposes no constraint on the equilibria explored below, and we do not invoke it further.

	\section{Dimensionless formulation and the compactness parameter}
	\label{sec:dimensionless}

	It is convenient to recast the equilibrium equations in dimensionless form. I first introduce the physical parameters and characteristic scales that label a configuration (Section~\ref{sec:dimless_params}), and then assemble the dimensionless boundary-value equations that are integrated numerically (Section~\ref{sec:dimless_construction}); the inversion of the Walecka relation, required only when the Fermi momentum approaches the particle mass, is given in Appendix~\ref{app:walecka_inversion}. A complete glossary of the symbols and rescaled quantities is collected in Appendix~\ref{app:notation}.

	\subsection{Physical parameters and characteristic scales}
	\label{sec:dimless_params}

	I introduce the dimensionless coordinate $x=r/r_0$, where $r_0$ is a free length scale fixed by the boundary conditions, and normalized scalar-density amplitudes
	\begin{equation}
		\eta_{s,i}(x)=\sqrt{n_{s,i}(r)/n_0}\,,
		\label{eq:eta_s_def}
	\end{equation}
	with $n_0$ the central number density of species 1. The nonrelativistic limit is obtained by the replacement $\eta_{s,i}\to \eta_i = \sqrt{n_i/n_0}$, where $n_i$ are the particle number densities. A normalized gravitational potential $\varphi(x)$ completes the variable set.

	The natural energy and density scales of the problem are set by the Bohm kinetic energy per unit mass evaluated at the scale $r_0$:
	\begin{equation}
		\Phi_0 = \frac{\lamB\hbar^2}{m_1^2\,r_0^2}\,,\qquad
		\psi_0^2 = \frac{4\pi\lamB\hbar^2}{\gphi^2\,m_1\,r_0^4}\,.
		\label{eq:scales}
	\end{equation}
	Here $\Phi_0$ has dimensions of specific energy (energy per unit mass) and sets the gravitational potential scale: the dimensionless potential is $\varphi(x)\equiv\Phi(r)/\Phi_0$, where $\Phi(r)$ is the Newtonian gravitational potential.
	The quantity $\psi_0^2$ is the corresponding number-density scale, obtained by balancing the Bohm quantum pressure against the Yukawa attraction, so that $n_0=\psi_0^2\,\eta_{s,1}^2(0)$.
	The central boundary conditions are $\eta_{s,1}(0)=1$ and $\eta_{s,2}(0)=f$, where $f\in(0,1]$ is the central density fraction of the second species.
	The dimensionless pressure parameters are
	\begin{equation}
		\sigma_i \equiv \frac{m_1\,r_0^2}{\lamB\hbar^2}\,\frac{K_{p,i}\gamma_{p,i}}{\gamma_{p,i}-1}\,\psi_0^{2(\gamma_{p,i}-1)}\,,
		\qquad i\in\{1,2\}\,,
		\label{eq:sigma_def}
	\end{equation}
	with $\sigma_i=0$ recovering the pure Schrödinger--Poisson limit. The dimensionless mediator-mass parameter is $\xi \equiv \kappaphi\,r_0 = \mphi\,c\,r_0/\hbar$; setting $\xi=0$ recovers the massless-mediator limit.
	Finally, the species mass ratio $q \equiv m_1/m_2\geq 1$ measures how much lighter the second species is than the first; the single-species configuration is recovered in the limit $q\to 1$ with $f\to 0$. Its role in the rescaled equations is discussed in Section~\ref{sec:dimless_construction}.

	The key connection to physical observables is made through the physical compactness parameter,
	\begin{equation}
		\kappa = \frac{GM}{R_T c^2}\,,
		\label{eq:kappa_physical}
	\end{equation}
	which measures the ratio of the gravitational potential energy at the surface to the rest-mass energy. In the dimensionless formulation, the related parameter is
	\begin{equation}
		\alpha \equiv \frac{\Phi_0}{c^2} = \frac{\lamB\hbar^2}{m_1^2\,r_0^2\,c^2}\,,
		\label{eq:alpha_def}
	\end{equation}
	which measures the ratio of the characteristic gravitational potential energy to the rest-mass energy. These parameters are related through
	\begin{equation}
		\kappa = \alpha\,\frac{M_{\mathrm{dim}}}{x_T}\,,
		\label{eq:kappa_alpha_rel}
	\end{equation}
	where $M_{\mathrm{dim}}$ and $x_T$ are the dimensionless mass and radius. The physical compactness $\kappa$ is the primary observable, while $\alpha$ parameterises the dimensionless equations.

	\subsection{Construction of the dimensionless boundary-value equations}
	\label{sec:dimless_construction}

	The dimensionless GR system is derived by non-dimensionalising the three governing physical equations: the modified TOV equation~\eqref{eq:TOV_bohm} for each species, and the nonlinear Klein--Gordon equation~\eqref{eq:KG_nuclear} for the mediator.

	\medskip\noindent\emph{Matter sector.}---The closure relation~\eqref{eq:GR_closure} expresses the polytropic pseudoenthalpy $U_{P,i}$ as a function of $n_i$, the metric potential $\nu$, and the Bohm potential $Q_i^{\mathrm{GR}}$; taking its gradient reproduces Eq.~\eqref{eq:TOV_bohm} identically, so the two equations may be used interchangeably.
	Substituting the scalar-density amplitude $\eta_{s,i}=\sqrt{n_{s,i}/n_0}$ into the curved-spacetime Bohm potential~\eqref{eq:bohm_GR}, one converts the first-order force-balance equation into a second-order equation for $\eta_{s,i}$ in which the covariant Laplacian~\eqref{eq:cov_laplacian} appears explicitly on the left-hand side.
	The metric-potential gradient $\mathrm{d}\nu/\mathrm{d}r$ is supplied by the mass equation~\eqref{eq:mass_eq}, which tracks the enclosed mass $m(r)$, and the mass equation itself becomes a quadrature for the dimensionless enclosed mass $\mathcal{M}_c(x)$.
	Rescaling to $x=r/r_0$ and the characteristic scales~\eqref{eq:scales}, and introducing $\alpha$~\eqref{eq:alpha_def} together with the dimensionless metric function $A(x)\equiv 1-2\alpha\,\mathcal{M}_c(x)/x$ (the dimensionless form of $A(r)$ defined in Eq.~\eqref{eq:metric_function}, obtained by setting $m=M_0\mathcal{M}_c$ and $r=r_0 x$; identifying $\lambda'=-A'/(2A)$ then casts the covariant Laplacian~\eqref{eq:box_to_cov} into the form $A[\phi''+(2/r+\nu'+A'/(2A))\phi']$), gives
	\begin{align}
	    \eta_{s,1}''+\frac{2}{x}\eta_{s,1}'
	    &=\frac{1}{A(x)}\bigl[2\beta_1 v_c\,\eta_{s,1}\,\varphi + 2\sigma_1\,\eta_1^{2\gamma_{p,1}-1} - 2\mathcal{E}_1\,\eta_{s,1}\bigr] \nonumber\\
	    &\quad -\bigl[\nu'_d+\tfrac{A'}{2A}\bigr]\eta_{s,1}'\,,
	    \label{eq:gr_eta1}\\
	    \eta_{s,2}''+\frac{2}{x}\eta_{s,2}'
	    &=\frac{1}{qA(x)}\bigl[\tfrac{2\beta_2 v_c}{q}\,\eta_{s,2}\,\varphi + 2\sigma_2\,\eta_2^{2\gamma_{p,2}-1} - 2\mathcal{E}_2\,\eta_{s,2}\bigr] \nonumber\\
	    &\quad -\bigl[\nu'_d+\tfrac{A'}{2A}\bigr]\eta_{s,2}'\,.
	    \label{eq:gr_eta2}
	\end{align}
	A remark on notation is warranted. Equations~\eqref{eq:gr_eta1}--\eqref{eq:gr_eta2} deliberately mix two distinct density amplitudes: the Laplacian, Yukawa-coupling, and eigenvalue terms involve the scalar-density amplitude $\eta_{s,i}=\sqrt{n_{s,i}/n_0}$, because the Bohm potential and the Klein--Gordon source are constructed from $n_{s,i}$, whereas the polytropic pressure term involves the number-density amplitude $\eta_i=\sqrt{n_i/n_0}$, because the equation of state $P_i=K_{p,i}\,n_i^{\gamma_{p,i}}$ is defined in terms of the particle number density $n_i$.
	In the nonrelativistic limit, $\eta_{s,i}\to\eta_i$ everywhere and the distinction vanishes; at nuclear densities, the two amplitudes are related through the Walecka momentum integral~\eqref{eq:ns_walecka}, which depends on the effective mass $m_i^*$ and must be evaluated self-consistently at each radial point.
	The inversion of this relation, by which the number-density amplitude $\eta_i$ is recovered from the scalar-density amplitude $\eta_{s,i}$, is given in Appendix~\ref{app:walecka_inversion}.

	\medskip\noindent\emph{Mediator sector.}---Non-dimensionalising the full nonlinear Klein--Gordon equation~\eqref{eq:KG_nuclear} with $\phi=\phi_0\,\varphi$ (where $\phi_0$ is the characteristic field amplitude set by the source strength), $n_{s,i}=n_0\,\eta_{s,i}^2$, and the same coordinate rescaling yields the general dimensionless mediator equation
	\begin{multline}
	    \varphi''+\!\left(\frac{2}{x}+\nu'_d+\frac{A'}{2A}\right)\!\varphi'
	    = \frac{1}{A}\biggl[\frac{4\pi}{v_c}\!\left(\beta_1\eta_{s,1}^2+\frac{\beta_2\eta_{s,2}^2}{q^2}\right)\\
	    +\;\xi^2\,\varphi
	    +\tilde\kappa_3\,\varphi^2+\tilde\lambda_\phi\,\varphi^3\biggr],
	    \label{eq:gr_phi_general}
	\end{multline}
	where the dimensionless self-coupling parameters are
	\begin{equation}
	    \tilde\kappa_3 \equiv \frac{\kappa_3\,\phi_0\,r_0^2}{2}\,,\qquad
	    \tilde\lambda_\phi \equiv \frac{\lambda_\phi\,\phi_0^2\,r_0^2}{6}\,,
	    \label{eq:dimless_selfcouplings}
	\end{equation}
	and $\eta_{s,i}^2\equiv n_{s,i}/n_0$ are the dimensionless scalar densities.

	Equation~\eqref{eq:gr_phi_general} retains the full covariant Laplacian on a static, spherically symmetric background: the additional friction term $(\nu'_d+A'/(2A))\varphi'$ and the overall $1/A$ factor on the right-hand side are both formally $\mathcal{O}(\alpha)$ corrections to the flat-space Klein--Gordon operator.
	Although these contributions can be locally comparable to the source term at intermediate radii, their integrated effect on global observables remains modest: numerical tests show that the dimensionless mass $M_{\mathrm{dim}}$, the truncation radius $x_T$, and the dark-matter fraction $\kappa$ shift by less than $1\%$ for $\alpha\leq 0.25$, rising to approximately $2\%$ at $\alpha=0.30$.
	The smallness of the net correction is understood by noting that the mediator field is slaved to the fermion density profile and feeds back only through the Yukawa coupling, so that the covariant terms enter the structure at second order in $\alpha$.
	Equation~\eqref{eq:gr_phi_general} together with Eqs~\eqref{eq:gr_eta1}--\eqref{eq:gr_eta2} constitutes the most general dimensionless boundary-value problem of this work.
	For the ultralight baseline explored in Section~\ref{sec:results}, the self-couplings vanish ($\tilde\kappa_3=\tilde\lambda_\phi=0$) and $\eta_{s,i}\to\eta_i$, reducing the mediator equation to
	\begin{equation}
	    \varphi''+\!\left(\frac{2}{x}+\nu'_d+\frac{A'}{2A}\right)\!\varphi'
	    =\frac{1}{A}\!\left[\frac{4\pi}{v_c}\!\left(\beta_1\eta_1^2+\frac{\beta_2\eta_2^2}{q^2}\right)+\xi^2\,\varphi\right].
	    \label{eq:gr_phi}
	\end{equation}
	The numerical method of Section~\ref{sec:numerics} solves the general equation~\eqref{eq:gr_phi_general}, with Eq.~\eqref{eq:gr_phi} recovered in the appropriate limit.

	In both regimes, $A(x)\equiv 1-2\alpha\,\mathcal{M}_c(x)/x$ is the dimensionless metric function, $\nu'_d=\alpha\,\mathcal{M}_c/(x^2 A)$ the metric-potential gradient, and the covariant-Laplacian correction is
	\begin{equation}
	    \nu'_d+\frac{A'}{2A} = \frac{\alpha}{x^2 A}\!\left[2\mathcal{M}_c - 4\pi x^3\!\left(\eta_{s,1}^2+\frac{\eta_{s,2}^2}{q^2}\right)\right].
	    \label{eq:cov_correction}
	\end{equation}
	At $\alpha=0$ the Newtonian system is recovered exactly.

	The dimensionless mass and physical scales are
	\begin{equation}
		M_{\mathrm{dim}}(\alpha) = 4\pi\int_0^\infty \!\left(\eta_{s,1}^2+\frac{\eta_{s,2}^2}{q^2}\right) x^2\,\mathrm{d}x\,,
		\label{eq:Mdim}
	\end{equation}
	\begin{equation}
		M = \frac{\lamB\hbar^2}{G m_1^2 r_0}\,M_{\mathrm{dim}}(\alpha)\,,\qquad
		\RT = \xT(\alpha)\,r_0\,,
		\label{eq:MR_physical}
	\end{equation}
	with mass--radius relation
	\begin{equation}
		M\,\RT = \frac{\lamB\hbar^2}{G m_1^2}\,M_{\mathrm{dim}}(\alpha)\,\xT(\alpha)\,.
		\label{eq:MR_GR}
	\end{equation}

	The mass ratio $q$ pervades the equations assembled above because species~2 couples to the gravitational potential in proportion to its mass $m_2=m_1/q$, while its Bohm quantum pressure ($Q_i^{\mathrm{GR}}\propto\hbar^2/m_i$) is correspondingly stronger by the same factor. Each species-2 contribution therefore carries an explicit factor of $1/q$ in the density equation~\eqref{eq:gr_eta2}, and of $1/q^2$ in the source term of the mediator equation~\eqref{eq:gr_phi_general} and in the mass integral~\eqref{eq:Mdim}.

	\section{Numerical method}
	\label{sec:numerics}

	The complete dimensionless boundary-value problem, Eqs~\eqref{eq:gr_eta1}--\eqref{eq:gr_phi_general}, is solved numerically in the form described below.
	The formulation retains the scalar-density amplitudes $\eta_{s,i}=\sqrt{n_{s,i}/n_0}$ and all self-coupling parameters throughout, so that the various limiting cases emerge naturally from the general solver rather than being imposed \emph{a priori}.
	The system is recast as a first-order boundary-value problem (BVP) by introducing auxiliary variables $\zeta_{s,i}=x^2\,\mathrm{d}\eta_{s,i}/\mathrm{d}x$ and $\varpi=x^2\,\mathrm{d}\varphi/\mathrm{d}x$, yielding a seven-component system for the two-species case ($\eta_{s,1}$, $\zeta_{s,1}$, $\eta_{s,2}$, $\zeta_{s,2}$, $\mathcal{M}_c$, $\varphi$, $\varpi$) with eigenvalues ($\mathcal{E}_1$, $\mathcal{E}_2$, $v_c$), or a five-component system for single-species configurations.
	The state variables are the \emph{scalar-density} amplitudes $\eta_{s,i}=\sqrt{n_{s,i}/n_0}$; the corresponding \emph{number-density} amplitudes $\eta_i=\sqrt{n_i/n_0}$, needed for the polytropic pressure terms, are obtained at each grid point by inverting the Walecka relation~\eqref{eq:etas_eta_inversion}.
	The mediator equation, in the first-order representation, reads
	\begin{multline}
		\frac{\mathrm{d}\varpi}{\mathrm{d}x} = -\!\left(\nu'_d+\frac{A'}{2A}\right)\!\varpi \\
		+\;\frac{x^2}{A}\!\left[\frac{4\pi}{v_c}\!\left(\beta_1\eta_{s,1}^2+\frac{\beta_2\eta_{s,2}^2}{q^2}\right)+\xi^2\,\varphi +\tilde\kappa_3\,\varphi^2+\tilde\lambda_\phi\,\varphi^3\right],
		\label{eq:dchi_firstorder}
	\end{multline}
	which retains the cubic and quartic vertices explicitly.
	In the single-species limit ($f=0$ or $q=1$), the $\eta_{s,2}$ equation is suppressed and the BVP reduces to five components.
	The boundary conditions are
	\begin{equation}
		\begin{aligned}
			\eta_{s,1}(0)&=1,\quad \eta_{s,1}'(0)=0,\quad
			\eta_{s,2}(0)=f,\quad \eta_{s,2}'(0)=0,\\
			\varphi(0)&=1,\quad \varphi'(0)=0,\quad \mathcal{M}_c(0)=0\,,\\
			\eta_{s,1}(x_{\max})&=0,\quad \eta_{s,2}(x_{\max})=0,\quad \varphi(x_{\max})=0\,,
		\end{aligned}
		\label{eq:BC_GR}
	\end{equation}
	with eigenvalues $\mathcal{E}_1$, $\mathcal{E}_2$ and $v_c$ determined by the solver.
	Throughout the dimensionless formulation, $\mathcal{E}_i$ denotes the (negative, dimensionless) eigenvalue of species~$i$, not to be confused with the energy density $\varepsilon_i$ of the structure equations.
	In the nonrelativistic regime, $\eta_{s,i}\to\eta_i$ everywhere and the scalar densities reduce to the number densities; the boundary conditions and eigenvalue structure remain identical.

	The problem is, at its core, a nonlinear eigenvalue problem: for a given set of physical parameters ($\alpha$, $q$, $f$, $\sigma_i$, $\beta_i$, $\xi$, $\tilde\kappa_3$, $\tilde\lambda_\phi$), the solver must determine the eigenvalues $\mathcal{E}_1$, $\mathcal{E}_2$ and $v_c$ such that the density amplitudes and the mediator field simultaneously satisfy regularity at the origin and vanishing at the truncation radius. The eigenvalue $\mathcal{E}_i$ plays the role of a dimensionless chemical potential for species~$i$, encoding the balance between gravitational binding, Yukawa attraction, quantum pressure and polytropic degeneracy pressure; $v_c$ sets the depth of the central gravitational well. No closed-form relation exists among these three quantities, so they must be obtained as part of the global solution. For single-species configurations ($f=0$), only $\mathcal{E}_1$ and $v_c$ appear.

	At each collocation point, the polytropic pressure terms require the number-density amplitude~$\eta_i$, obtained from the BVP state variable~$\eta_{s,i}$ by Newton iteration on the Walecka relation~\eqref{eq:etas_eta_inversion} (typically 3--5 iterations to machine precision); when $\chi_0=0$, the inversion reduces to the identity $\eta_i=\eta_{s,i}$.

	The BVP is solved by three-stage Lobatto collocation~\citep{2001ACMTMS..27..299K} as implemented in \textsc{SciPy}~\citep{2020NatMe..17..261V}, with $x_{\max}=30$, 600--900 mesh points, and relative residual tolerance of $5\times 10^{-7}$.
	At $\alpha=0$ and $\tilde\kappa_3=\tilde\lambda_\phi=0$, the code recovers the Newtonian Schr\"odinger--Poisson invariants ($M_{\mathrm{dim}}(0)=3.883$, $\xT(0)=2.575$, $M_{\mathrm{dim}}\,\xT=10.00$) within $0.5\%$ of benchmarked values \citep{1998CQGra..15.2733M,2011PhRvD..84d3531C,2004PhRvD..69l4033G}.
	The continuation strategy proceeds by gradually increasing $\alpha$ from zero, using each converged solution as the initial guess for the next step, which proves essential for attaining stable convergence at high compactness ($\alpha\gtrsim 0.15$).

	\section{Results}
	\label{sec:results}

\begin{table}
\centering
\caption{Dimensionless invariants for the single-species ground state as a function of the compactness parameter $\alpha$. The physical compactness is $\kappa=\alpha M_{\mathrm{dim}}/\xT$, and $\Delta(M_{\mathrm{dim}}\,\xT)$ denotes the fractional change of the mass--radius product relative to its $\alpha=0$ value. At $\alpha=0$ the established Newtonian benchmarks are recovered.}
\label{tab:gr_invariants}
\begin{tabular}{@{}cccccc@{}}
	\toprule
	$\alpha$ & $M_{\mathrm{dim}}$ & $\xT$ & $M_{\mathrm{dim}}\,\xT$ & $\kappa$ & $\Delta(M_{\mathrm{dim}}\,\xT)$ \\
	\midrule
	0      & 3.883 & 2.575 & 10.00 & 0 & $0\%$ \\
	0.001  & 3.869 & 2.575 &  9.96 & $1.5\times 10^{-3}$ & $-0.4\%$ \\
	0.005  & 3.814 & 2.554 &  9.74 & $7.5\times 10^{-3}$ & $-2.6\%$ \\
	0.01   & 3.747 & 2.532 &  9.49 & $1.48\times 10^{-2}$ & $-5.1\%$ \\
	0.02   & 3.616 & 2.489 &  9.00 & $2.91\times 10^{-2}$ & $-10.0\%$ \\
	0.05   & 3.253 & 2.361 &  7.68 & $6.89\times 10^{-2}$ & $-23.2\%$ \\
	0.10   & 2.734 & 2.189 &  5.98 & 0.125 & $-40.2\%$ \\
	0.15   & 2.306 & 2.039 &  4.70 & 0.170 & $-53.0\%$ \\
	0.20   & 1.954 & 1.931 &  3.77 & 0.202 & $-62.3\%$ \\
	0.25   & 1.666 & 1.845 &  3.07 & 0.226 & $-69.3\%$ \\
	0.30   & 1.433 & 1.803 &  2.58 & 0.238 & $-74.2\%$ \\
	\bottomrule
\end{tabular}
\end{table}

	The numerical results presented in this section establish the dimensionless structure of the solutions, from the pure-Bohm baseline through to configurations of substantial compactness and the nuclear-density regime.

	\subsection{Pure-Bohm baseline: compactness and coupling-ratio dependence}

	Table~\ref{tab:gr_invariants} presents the dimensionless mass, the $99\%$-mass radius, and their product as a function of $\alpha$ for the single-species Schrödinger--Poisson baseline ($\sigma_i=0$, $f=0$, $\xi=0$).

	Both $M_{\mathrm{dim}}$ and the mass--radius product decrease monotonically with increasing $\alpha$, a result that is physically intuitive: in the full theory pressure itself contributes to the gravitational source, thereby enhancing the effective confinement, much as adding weight to a spring compresses it further than the initial load alone would suggest.
	At $\alpha=0.01$ ($\kappa\approx 0.015$), the mass--radius product is $5\%$ below its $\alpha=0$ value; at $\alpha=0.1$ ($\kappa\approx 0.125$), the reduction reaches $40\%$, confirming that the compact configurations of primary observational interest lie firmly in the regime where the full general-relativistic framework is required.

	\begin{figure}
	\centering
	\includegraphics[width=\columnwidth]{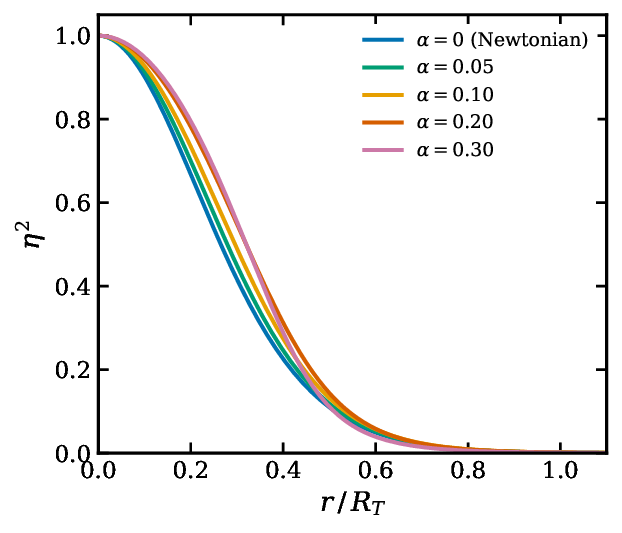}
	\caption{Normalized density profiles $\eta^2$ as a function of $r/R_T$ for the single-species ground state at $\alpha=0$ (Newtonian), $0.05$, $0.10$, $0.20$ and $0.30$. Here $\eta\equiv\sqrt{n/n_0}$ is the number-density amplitude; in the nonrelativistic regime shown, it coincides with the scalar-density amplitude $\eta_s$. Increasing compactness concentrates the density towards the center.}
	\label{fig:profiles}
\end{figure}

	To isolate the effect of the coupling ratio, Table~\ref{tab:beta_invariants} surveys the same pure-Bohm baseline as Table~\ref{tab:gr_invariants}, but now varying $\beta_i$ (subgravitational at $0.5$, comparable to gravity at $1$, supergravitational at $5$ and $10$) jointly with $\alpha$.
	Reference physical radii $R_T$ are quoted for $m_1=6\times10^{-14}\,\eV$ and $M=1\,\Msun$; for other choices, $R_T\propto M_{\mathrm{dim}}/(\beta_i\,m_1^2\,M)$ via Eq.~\eqref{eq:MR_physical}.

	As $\beta_i$ increases, the Yukawa attraction strengthens relative to gravity, drawing the configuration inward and reducing both $M_{\mathrm{dim}}$ and $\xT$.
	The Bohm quantum pressure ($Q_i^{\mathrm{GR}}\propto\hbar^2/m_i$) opposes this contraction and sets the equilibrium radius; the density profile at each $\beta_i$ reflects the interplay between the deepening Yukawa well and the stiffening quantum pressure.
	The fractional GR reduction in $M_{\mathrm{dim}}\,\xT$ between $\alpha=0$ and $\alpha=0.30$ depends strongly on $\beta_i$: $15$--$27\%$ at $\beta_i=5$--$10$, rising to $74\%$ at $\beta_i=1$, with $\kappa$ reaching $\approx 0.24$ for the most compact configurations.
	For subgravitational coupling $\beta_i=0.5$ the ground-state branch terminates at $\alpha\approx 0.20$, the covariant Laplacian correction destabilising the compact solution.
	Sustaining still higher compactness ($\kappa\gtrsim 0.25$) requires polytropic degeneracy pressure ($\sigma\neq 0$), as realised in Table~\ref{tab:physical}.

	\begin{table}
	\centering
	\caption{Dimensionless invariants $M_{\mathrm{dim}}(\alpha,\beta_i)$ and $\xT(\alpha,\beta_i)$ as a function of coupling ratio $\beta_i$ and compactness parameter $\alpha$, together with the gravitational compactness $\kappa=GM/(R_T c^2)=\alpha\,M_{\mathrm{dim}}/\xT$ and the total radius $R_T$ evaluated for reference parameters $m_1=6\times10^{-14}\,\eV$, $M=1\,\Msun$. All models use pure quantum pressure ($\sigma=0$).}
	\label{tab:beta_invariants}
	\begin{tabular}{@{}ccccccc@{}}
		\toprule
		$\beta_i$ & $\alpha$ & $M_{\mathrm{dim}}$ & $\xT$ & $M_{\mathrm{dim}}\,\xT$ & $\kappa$ & $R_T\;[\mathrm{km}]$ \\
		\midrule
		0.5 & 0    & 10.98 & 3.63 & 39.86 & 0     & $3.2\times10^{7}$ \\
		0.5 & 0.10 &  5.53 & 2.72 & 15.04 & 0.203 & $1.2\times10^{7}$ \\
		0.5 & 0.20 &  3.15 & 2.87 &  9.01 & 0.219 & $7.3\times10^{6}$ \\
		0.5 & 0.30 &  ---  & ---  &  ---  & ---   & ---               \\[3pt]
		1.0 & 0    &  3.88 & 2.58 & 10.00 & 0     & $8.1\times10^{6}$ \\
		1.0 & 0.10 &  2.73 & 2.19 &  5.98 & 0.125 & $4.9\times10^{6}$ \\
		1.0 & 0.20 &  1.95 & 1.93 &  3.77 & 0.202 & $3.1\times10^{6}$ \\
		1.0 & 0.30 &  1.43 & 1.80 &  2.58 & 0.238 & $2.1\times10^{6}$ \\[3pt]
		2.0 & 0    &  1.37 & 1.82 &  2.50 & 0     & $2.0\times10^{6}$ \\
		2.0 & 0.10 &  1.15 & 1.67 &  1.93 & 0.069 & $1.6\times10^{6}$ \\
		2.0 & 0.20 &  0.967 & 1.55 & 1.49 & 0.125 & $1.2\times10^{6}$ \\
		2.0 & 0.30 &  0.815 & 1.46 & 1.19 & 0.168 & $9.7\times10^{5}$ \\[3pt]
		5.0 & 0    &  0.347 & 1.16 & 0.402 & 0     & $3.3\times10^{5}$ \\
		5.0 & 0.10 &  0.323 & 1.12 & 0.360 & 0.029 & $2.9\times10^{5}$ \\
		5.0 & 0.20 &  0.301 & 1.07 & 0.323 & 0.056 & $2.6\times10^{5}$ \\
		5.0 & 0.30 &  0.281 & 1.04 & 0.292 & 0.081 & $2.4\times10^{5}$ \\[3pt]
		10  & 0    &  0.123 & 0.82 & 0.100 & 0     & $8.2\times10^{4}$ \\
		10  & 0.10 &  0.118 & 0.81 & 0.095 & 0.015 & $7.8\times10^{4}$ \\
		10  & 0.20 &  0.114 & 0.79 & 0.091 & 0.029 & $7.4\times10^{4}$ \\
		10  & 0.30 &  0.110 & 0.77 & 0.085 & 0.043 & $6.9\times10^{4}$ \\
		\bottomrule
	\end{tabular}
	\end{table}

	\subsection{Density profiles and potential balance}

	\begin{figure}
	\centering
	\includegraphics[width=\columnwidth]{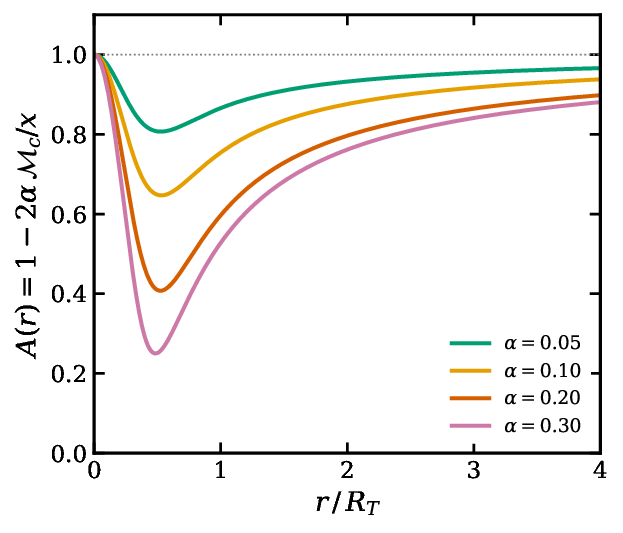}
	\caption{Metric function $A(x)=1-2\alpha\,\mathcal{M}_c(x)/x$ for the single-species ground state as a function of $r/R_T$ ($\sigma=0$) at selected values of the compactness parameter $\alpha$. The central minimum deepens monotonically with increasing $\alpha$, signalling stronger spacetime curvature near the stellar center.}
	\label{fig:potential_balance}
\end{figure}

\begin{figure}
		\centering
		\includegraphics[width=\columnwidth]{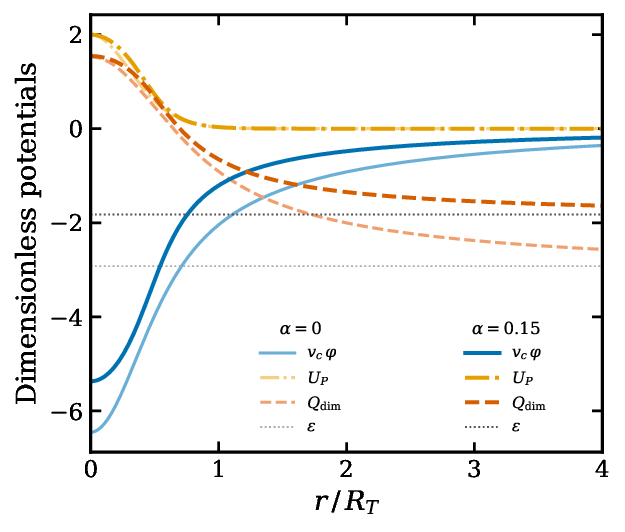}
		\caption{Polytropic potential balance for the single-species ground state with $\gamma_p=5/3$ and $\sigma=2$, comparing the Newtonian limit $\alpha=0$ (light curves) with the GR case $\alpha=0.15$ (dark curves). Blue solid: gravitational potential $v_c\varphi$; orange dot-dashed: polytropic pseudoenthalpy $U_P$; vermillion dashed: Bohm diagnostic $Q_{\mathrm{dim}}$; grey dotted: eigenvalue $\mathcal{E}$.}
		\label{fig:polytropic_balance}
	\end{figure}

	The results presented in this subsection are obtained for the single-species Schr\"odinger--Poisson ground state with pure Bohm quantum pressure ($\sigma=0$, $f=0$, $\beta_i=1$, $\tilde\kappa_3=\tilde\lambda_\phi=0$, $\xi=0$), the same baseline model as Table~\ref{tab:gr_invariants}. Figure~\ref{fig:profiles} shows the normalized density profile $\eta^2(x)$ at several values of $\alpha$.
	As $\alpha$ increases, the profile becomes appreciably more centrally concentrated: increasing compactness draws mass inward and steepens the density gradient, rather as deepening a finite square well confines the ground-state wave function ever more tightly.
	The half-mass radius decreases from approximately $1.2$ at $\alpha=0$ to approximately $1.0$ at $\alpha=0.2$.
	Figure~\ref{fig:potential_balance} shows the metric function $A(x)=1-2\alpha\,\mathcal{M}_c(x)/x$, whose central minimum deepens monotonically with $\alpha$, confirming the strengthening of spacetime curvature as the compactness increases.
	When a polytropic pseudoenthalpy is included ($\gamma_p=5/3$, $\sigma=2$), degeneracy-pressure support supplements the Bohm quantum pressure, and the eigenvalue rises from $\mathcal{E}=-2.922$ at $\alpha=0$ to $\mathcal{E}=-1.825$ at $\alpha=0.15$ (Fig.~\ref{fig:polytropic_balance}), the well becoming shallower as the increasing compactness lowers the dimensionless binding.

	\subsection{Two-species system and coupling-ratio dependence}

\begin{figure}
		\centering
		\includegraphics[width=0.95\columnwidth]{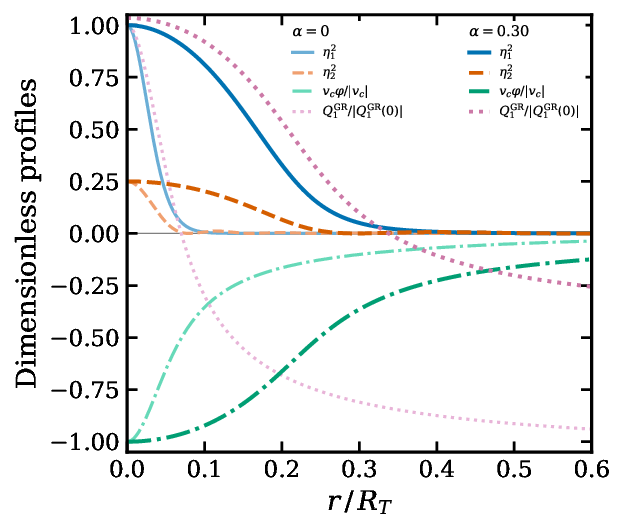}
		\caption{Two-species dimensionless profiles as a function of $r/R_T$ for $q=10$, $f=0.5$, $\xi=0$. Light curves: Newtonian limit ($\alpha=0$); dark curves: strongly relativistic regime ($\alpha=0.30$). Blue solid: $\eta_1^2$ (species~1); vermillion dashed: $\eta_2^2$ (species~2); green dot-dashed: normalized gravitational potential $v_c\varphi/|v_c|$; purple dotted: normalized Bohm quantum pressure $Q_1^{\mathrm{GR}}/|Q_1^{\mathrm{GR}}(0)|$.}
		\label{fig:two_species}
	\end{figure}

	For the two-species system ($q=10$, $f=0.5$, $\sigma_i=0$, $\xi=0$), increasing the compactness parameter further modifies the spatial segregation between the two components (Fig.~\ref{fig:two_species}).
	At $\alpha=0.30$, the density profiles are compressed substantially inward: species~1 is concentrated within a smaller effective radius, and the gravitational potential well deepens markedly, reflecting the enhanced self-gravity characteristic of high-compactness configurations.
	Because species~2, with mass $m_2=m_1/q$, couples to the gravitational potential $q$ times more weakly than species~1 while its Bohm quantum pressure ($Q_i^{\mathrm{GR}}\propto\hbar^2/m_i$) is correspondingly enhanced, it extends to substantially larger radii, carrying only approximately $4\%$ of the total mass.

\begin{figure}
		\centering
		\includegraphics[width=0.95\columnwidth]{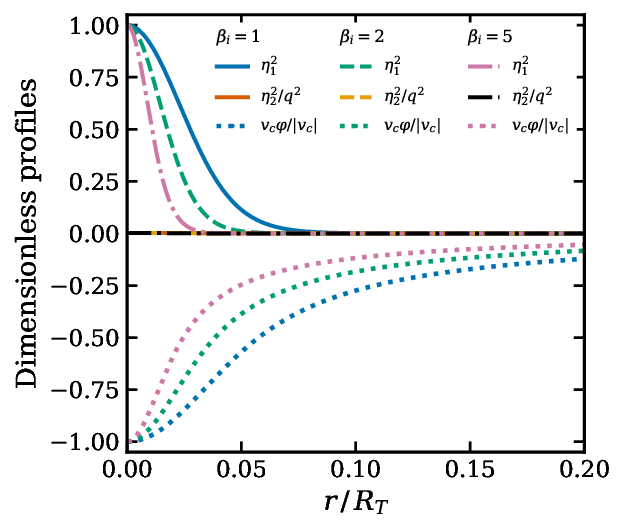}
		\caption{Two-species density profiles and normalized Yukawa mediator field ($q=10$, $f=0.5$, $\alpha=0.05$) at three coupling ratios $\beta_i=1$, $2$ and $5$. Species~1 ($\eta_1^2$, solid/dashed/dot-dashed), species~2 ($\eta_2^2/q^2$, same styles), and the normalized mediator potential $v_c\varphi/|v_c|$ (dotted) are shown for each $\beta_i$.}
		\label{fig:beta_2species}
	\end{figure}

	The dependence of the two-species density profiles and mediator field on $\beta_i$ is shown in Fig.~\ref{fig:beta_2species} for three representative coupling ratios.
	As $\beta_i$ increases, the Yukawa well deepens and the density profiles contract inward, confirming the enhanced confinement driven by the fifth force.

	\subsection{Self-coupling effects}
	\label{sec:selfcoupling}

\begin{figure}
		\centering
		\includegraphics[width=0.95\columnwidth]{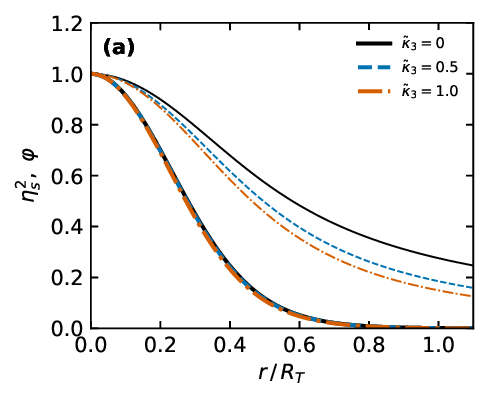}\\[4pt]
		\includegraphics[width=0.95\columnwidth]{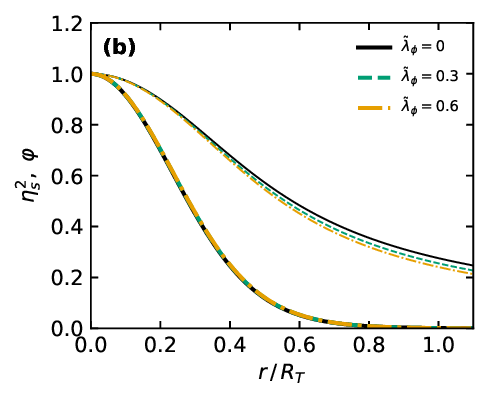}
		\caption{Effect of mediator self-couplings on the equilibrium profiles at $\alpha=0.05$, plotted against $r/R_T$. In each panel, thick curves show the density $\eta_s^2$ and thin curves the mediator potential $\varphi$. (a)~Varying cubic coupling $\tilde\kappa_3$ (with $\tilde\lambda_\phi=0$): increasing $\tilde\kappa_3$ steepens the central mediator profile and contracts the tidal radius. (b)~Varying quartic coupling $\tilde\lambda_\phi$ (with $\tilde\kappa_3=0$): increasing $\tilde\lambda_\phi$ provides a repulsive core that broadens the density distribution. The black curve is the ultralight baseline ($\tilde\kappa_3=\tilde\lambda_\phi=0$) in both panels.}
		\label{fig:selfcoupling}
	\end{figure}

	Having established the ultralight baseline, we now consider the effect of the self-coupling parameters $\tilde\kappa_3$ and $\tilde\lambda_\phi$ at fixed compactness $\alpha=0.05$.
	Figure~\ref{fig:selfcoupling} shows the density profiles $\eta^2$ (thick curves) and mediator field $\varphi$ (thin curves) for $\tilde\kappa_3 = 0$, $0.5$, $1.0$ (with $\tilde\lambda_\phi=0$) and for $\tilde\lambda_\phi = 0$, $0.3$, $0.6$ (with $\tilde\kappa_3=0$).
	The cubic coupling steepens the central profile and contracts the tidal radius, the $\varphi^3$ vertex shifting the effective mediator mass; the quartic coupling provides a repulsive core that broadens the density distribution, the $\varphi^4$ vertex furnishing a stabilising restoring force.

	The role of the self-couplings here differs qualitatively from that in the boson-star literature~\citep{1986PhRvL..57.2485C}.
	In a boson star, the scalar field \emph{is} the matter: the quartic self-interaction $\lambda|\phi|^4/4$ governs the equation of state, and the astrophysically relevant combination $\Lambda=\lambda\,M_{\mathrm{Pl}}^2/(4\pi\,m^2)$ can be enormous even for tiny bare couplings, owing to the Planck-mass amplification.
	In the present framework $\phi$ is the mediator of the Yukawa interaction between degenerate fermions, so its amplitude (sourced by fermion densities) is much smaller than the boson-star value at which $\phi$ must support the entire gravitational mass; the dimensionless combinations $\tilde\kappa_3$ and $\tilde\lambda_\phi$ accordingly scale with $\phi_0$ rather than $M_{\mathrm{Pl}}$.
	The perturbative unitarity bounds $\lambda_\phi\lesssim 4\pi$ and $\kappa_3\lesssim\mphi\sqrt{4\pi}$~\citep{1977PhRvL..38..883L,1977PhRvD..16.1519L,2018EPJC...78..748G} remain valid for the mediator sector; the values $\tilde\kappa_3=0$--$1.0$ and $\tilde\lambda_\phi=0$--$0.6$ adopted above correspond to order-unity perturbations, probing the regime in which self-couplings produce observable corrections without dominating the equilibrium structure.

	\subsection{Nuclear-density quantum dark fermion stars}
	\label{sec:nuclear_density}

	\begin{figure}
	\centering
	\includegraphics[width=0.95\columnwidth]{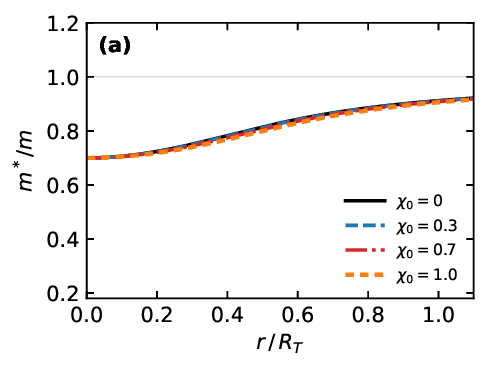}\\[4pt]
	\includegraphics[width=0.95\columnwidth]{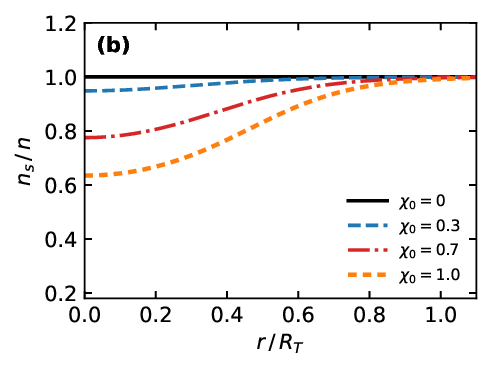}
	\caption{Walecka mean-field effects at $\alpha=0.05$, $\beta_i=1$, $\sigma=2$, $\gamma_p=5/3$, with effective-mass coupling $g_i\phi_0/(m_ic^2)=-0.3$. (a)~Effective-mass ratio $m^*/m=1+g_i\phi_0\,\varphi/(m_ic^2)$ as a function of $r/R_T$ for four values of the central relativity parameter $\chi_0=p_F/(mc)$. (b)~Scalar-density to number-density ratio $n_s/n=\eta_s^2/\eta^2$.}
	\label{fig:nuclear}
\end{figure}

	At densities where the Fermi momentum becomes comparable to the particle mass ($\chi_0\gtrsim 1$), the nonrelativistic replacement $n_{s,i}\to n_i$ must be abandoned and the general system~\eqref{eq:gr_eta1}--\eqref{eq:gr_phi_general} applies in full, with relativistic scalar densities, effective masses~\eqref{eq:mstar_def}, and the Walecka energy density and pressure~\eqref{eq:eps_RMF}--\eqref{eq:P_RMF}.

	\medskip\noindent\emph{(a) Boundary conditions.}---At the center, regularity requires
	\begin{equation}
		\begin{aligned}
			n_i'(0)&=0,\quad \phi'(0)=0,\quad m(0)=0,\\
			\nu'(0)&=0,\quad n_i(0)=n_{c,i},\quad \phi(0)=\phi_c\,,
		\end{aligned}
		\label{eq:BC_nuclear}
	\end{equation}
	where the central densities $n_{c,i}$ and central field amplitude $\phi_c$ are determined self-consistently.
	At the stellar surface, the densities and pressures vanish, and the metric functions match the exterior Schwarzschild solution.
	The eigenvalues consist of the two central densities and the central field amplitude.

	\medskip\noindent\emph{(b) Determination of the self-coupling parameters.}---The cubic and quartic couplings are bounded by perturbative unitarity: $\lambda_\phi\lesssim 4\pi$ and $\kappa_3\lesssim\mphi\sqrt{4\pi}$~\citep{1977PhRvL..38..883L,1977PhRvD..16.1519L,2018EPJC...78..748G}.
	Three complementary strategies constrain them:
	(i) Perturbative unitarity: tree-level $\phi\phi\to\phi\phi$ scattering must satisfy partial-wave unitarity~\citep{1977PhRvL..38..883L}.
	(ii) Self-consistency with saturation: if the dark sector exhibits nuclear saturation, the couplings are fixed by requiring a minimum in the binding energy per particle at some equilibrium density.
	(iii) Phenomenological fitting: once sufficient-quality gravitational-wave data become available, observed masses and radii may be matched to the theoretical mass--radius relation, as with neutron-star equations of state.

	\medskip\noindent\emph{(c) The ultrarelativistic $\gamma=4/3$ polytropic case.}---When the Fermi momentum of one or both species approaches the rest mass, the nonrelativistic equation of state $\gamma=5/3$ breaks down and must be replaced by $\gamma=4/3$, corresponding to ultrarelativistic degenerate fermions. The transition between the nonrelativistic and ultrarelativistic regimes occurs when the central Fermi momentum approaches $m_i c$, and the corresponding central density is $n_i\sim (m_i c)^3/(\pi^2\hbar^3)\sim 10^{42}\,\mathrm{cm}^{-3}$ for $m_1\sim 10^{-11}\,\eV$, far exceeding nuclear density. Nonetheless, configurations with high central density may produce modest deviations from the $\gamma=5/3$ law that, while falling short of true ultrarelativicity, are worth accounting for in precision equation-of-state studies.

	\medskip\noindent\emph{(d) Qualitative differences from the ultralight baseline.}---The effective-mass suppression ($m_i^*<m_i$) softens the equation of state, reducing the maximum mass and increasing the central density at the mass limit, in a manner broadly analogous to the Walecka nuclear model.
	The nonlinear self-coupling terms introduce a saturation mechanism: the quartic term eventually dominates and arrests field-amplitude growth, producing a finite equilibrium density rather than unbounded collapse.
	The interplay of these effects with the Bohm correction yields a richer mass--radius relation, with the possibility of multiple stable branches analogous to the neutron-star and quark-star branches in hadronic physics.

	Figure~\ref{fig:nuclear} illustrates the two principal Walecka effects.
	The effective-mass ratio $m^*/m$ (panel a) is suppressed to approximately $0.70$ at the center and recovers towards unity in the outer envelope where $\varphi\to 0$, with the curves varying only slightly with $\chi_0$ through the altered Walecka inversion.
	The scalar-to-number density ratio $n_s/n$ (panel b) departs from unity as $\chi_0$ grows, reaching approximately $0.63$ at the center for $\chi_0=1.0$, consistent with the Walecka estimate $n_s/n\sim 0.6$--$0.7$.
	The simultaneous deepening of the eigenvalue ($\mathcal{E}$ from $-2.62$ to $-3.30$) and increase in dimensionless mass ($M_{\mathrm{dim}}$ from $4.91$ to $6.23$) demonstrate that the Walecka mechanism produces quantitatively significant modifications well before the fully relativistic regime.
	The weak-coupling value $\alpha=0.05$ is adopted here because at stronger couplings the dimensionless mass becomes nearly independent of $\chi_0$, obscuring the structural changes.

	\begin{figure}
	\centering
	\includegraphics[width=\columnwidth]{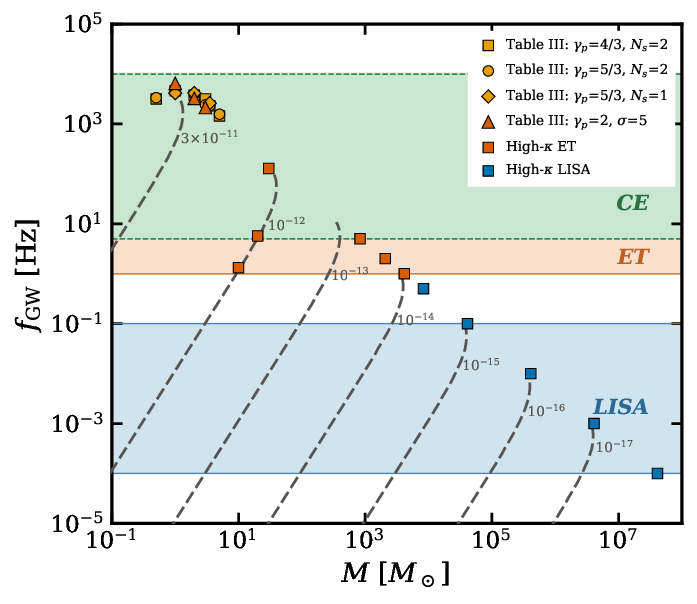}
	\caption{Gravitational-wave contact frequency $f_{\mathrm{GW}}$ (Eq.~\eqref{eq:fGW}) as a function of total mass $M$ for equal-mass dark fermion star binaries ($M_{\mathrm{tot}}=2M$). Shaded bands denote the approximate sensitivity windows of LISA ($10^{-4}$--$10^{-1}\,\mathrm{Hz}$ \cite{2017arXiv170200786A}), the Einstein Telescope (ET, $1$--$10^4\,\mathrm{Hz}$ \cite{2010CQGra..27s4002P}), and Cosmic Explorer (CE, $5$--$10^4\,\mathrm{Hz}$ \cite{2019BAAS...51g..35R}); the dashed green line marks the CE lower bound.
		Dashed grey curves show the self-consistent GR equilibrium sequences for seven dark-fermion masses $m_1$ (labelled, in eV), computed parametrically from the full GR solution grid for the two-fluid model with $\gamma_p=4/3$, $q=10$, $f=0.1$.
		Markers show all twenty-nine Table~\ref{tab:physical} configurations: orange squares ($\gamma_p=4/3$, $N_s=2$), orange circles ($\gamma_p=5/3$, $N_s=2$), orange diamonds ($\gamma_p=5/3$, $N_s=1$) and vermillion triangles ($\gamma_p=2$, $\sigma=5$) in the CE/ET band; vermillion squares for the intermediate-mass and high-compactness ET entries; blue squares for high-compactness LISA entries.}
	\label{fig:gw_predictions}
\end{figure}

	\section{Observational predictions}
	\label{sec:observations}

	The dimensionless equilibrium solutions of the preceding section acquire physical content once a particle mass $m_1$ is specified.
	This section translates the structural results for quantum dark fermion stars into observable signatures, gravitational-wave contact frequencies, mass--radius relations, and microlensing event parameters, with particular emphasis on the regime in which the predicted radii and compactnesses span from the neutron-star band ($R\sim 10$--$13\,\mathrm{km}$, $\kappa\sim 0.15$--$0.25$ \cite{2021ApJ...918L..28M}) up to the strongly relativistic regime ($\kappa\sim 0.34$, $R_T/R_S\simeq 1.5$), where the most compact configurations could masquerade as low-mass black holes in the lower mass-gap region.

\begin{figure}
	\centering
	\includegraphics[width=\columnwidth]{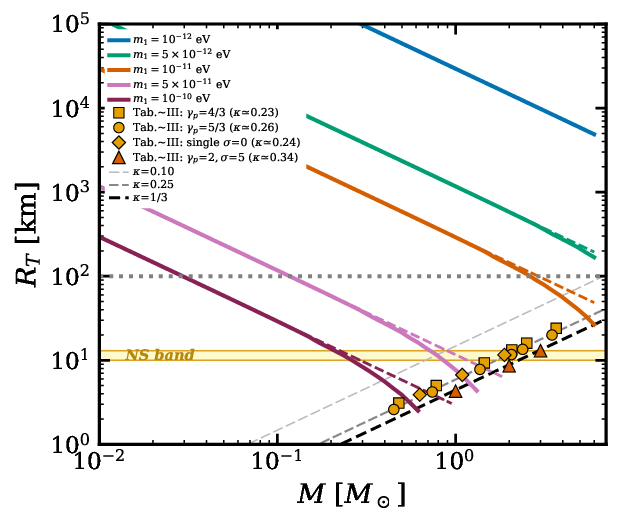}
	\caption{Mass--radius relation $R_T(M)$ for dark fermion stars. Solid curves show the full GR equilibrium sequences and dashed curves the Newtonian prediction for five particle masses ($m_1=10^{-12}$, $5\times10^{-12}$, $10^{-11}$, $5\times10^{-11}$, $10^{-10}\,\eV$, from top to bottom); a horizontal dotted line marks $R_T=100\,\mathrm{km}$. Diagonal grey dashed lines indicate constant-compactness loci at $\kappa=GM/(R_Tc^2)=0.10$, $0.25$, and the photon-sphere (light-ring) value $\kappa=1/3$. Orange markers show the fifteen CE/ET-band configurations from Table~\ref{tab:physical}: squares for $\gamma_p=4/3$ two-species ($\kappa\simeq0.23$), circles for $\gamma_p=5/3$ two-species ($\kappa\simeq0.26$), and diamonds for $\gamma_p=5/3$ single-species models ($\sigma=0$, $\kappa\simeq0.24$); vermillion triangles mark the three stiff, weak-Bohm configurations ($\gamma_p=2$, $\sigma=5$, $\kappa\simeq0.34$), which cross the light-ring line. The light shaded band at $R_T\simeq 10$--$13\,\mathrm{km}$ indicates the neutron-star radius range inferred from NICER X-ray timing and GW170817 tidal-deformability data \citep{2021ApJ...918L..28M,2018PhRvL.121p1101A}.}
	\label{fig:mass_radius}
\end{figure}

	\subsection{Physical scales and gravitational-wave predictions}

	Table~\ref{tab:physical} collects twenty-nine configurations spanning particle masses $m_1\sim 10^{-19}$--$10^{-10}\,\eV$ and total masses $M\sim 0.5$--$4\times10^{7}\,\Msun$, distributed across three groups: high-compactness CE/ET configurations near the neutron-star regime, intermediate-mass configurations bridging the weakly-to-strongly relativistic transition, and high-compactness ET/LISA configurations extending to supermassive scales.
	The contact gravitational-wave frequency for an equal-mass binary at touching separation is \citep{2007grav.book.....M}
	\begin{equation}
		f_{\mathrm{GW}} = \frac{1}{\pi}\sqrt{\frac{G M_{\mathrm{tot}}}{(2\RT)^3}}\,,
		\label{eq:fGW}
	\end{equation}
	with chirp mass $\mathcal{M}_{\mathrm{chirp}}=M_{\mathrm{tot}}/2^{6/5}$.

	\begin{table*}[ht!]
		\centering
	\caption{Quantum dark fermion star configurations spanning the LISA, Einstein
		Telescope (ET), and Cosmic Explorer (CE) gravitational-wave bands. All models
		include the Bohm quantum-pressure correction ($\lamB=1/9$). $N_s=2$ denotes
		two-species configurations ($f=0.1$, $\sigma=2$, mass ratio $q\equiv m_1/m_2=10$);
		$N_s=1$ denotes single-species models with pure quantum pressure ($\sigma=0$).
		For the single-species rows with $\sigma=0$ the polytropic term is absent,
		so $\gamma_p$ does not enter the equilibrium equations; the value $\gamma_p=5/3$
		shown reflects the physically appropriate nonrelativistic index at these
		densities ($p_F\ll m_1 c$). The single-species rows ($\sigma=5$)
			instead retain an active polytropic term, for which $\gamma_p=2$ sets the
			equation-of-state stiffness. $\kappa=GM/(R_Tc^2)$ is the physical compactness,
		and $\Delta(M\RT)$ the fractional reduction of the mass--radius product relative
		to the $\alpha\to 0$ limit.}
		\label{tab:physical}
		\begin{tabular}{@{}clccccccl@{}}
			\toprule
			$M$ [$\Msun$] & $m_1$ [$\eV$] & $\gamma_p$ & $N_s$ & $\kappa$ & $\RT$ & $f_{\mathrm{GW}}$ [Hz] & $\Delta(M\RT)$ & Band \\
			\midrule
			\multicolumn{9}{@{}l}{\textit{Two-species models, $\gamma_p=4/3$ -- CE/ET band}} \\[2pt]
			0.48 & $1\times 10^{-10}$ & $4/3$ & 2 & $0.23$ & $3.1\,\mathrm{km}$ & $7.5\times 10^{3}$ & $-72\%$ & CE/ET \\
			0.78 & $6.1\times 10^{-11}$ & $4/3$ & 2 & $0.23$ & $5.0\,\mathrm{km}$ & $4.6\times 10^{3}$ & $-72\%$ & CE/ET \\
			1.45 & $3.3\times 10^{-11}$ & $4/3$ & 2 & $0.23$ & $9.3\,\mathrm{km}$ & $2.5\times 10^{3}$ & $-72\%$ & CE/ET \\
			2.07 & $2.3\times 10^{-11}$ & $4/3$ & 2 & $0.23$ & $13.3\,\mathrm{km}$ & $1.7\times 10^{3}$ & $-72\%$ & CE/ET \\
			2.51 & $1.9\times 10^{-11}$ & $4/3$ & 2 & $0.23$ & $16\,\mathrm{km}$ & $1.4\times 10^{3}$ & $-72\%$ & CE/ET \\
			3.67 & $1.3\times 10^{-11}$ & $4/3$ & 2 & $0.23$ & $24\,\mathrm{km}$ & $9.8\times 10^{2}$ & $-72\%$ & CE/ET \\[4pt]
			\multicolumn{9}{@{}l}{\textit{Two-species models, $\gamma_p=5/3$ -- CE/ET band}} \\[2pt]
			0.45 & $1\times 10^{-10}$ & $5/3$ & 2 & $0.26$ & $2.6\,\mathrm{km}$ & $9.5\times 10^{3}$ & $-76\%$ & CE/ET \\
			0.74 & $6.1\times 10^{-11}$ & $5/3$ & 2 & $0.26$ & $4.2\,\mathrm{km}$ & $5.8\times 10^{3}$ & $-76\%$ & CE/ET \\
			1.37 & $3.3\times 10^{-11}$ & $5/3$ & 2 & $0.26$ & $7.8\,\mathrm{km}$ & $3.1\times 10^{3}$ & $-76\%$ & CE/ET \\
			2.06 & $2.2\times 10^{-11}$ & $5/3$ & 2 & $0.26$ & $11.7\,\mathrm{km}$ & $2.1\times 10^{3}$ & $-76\%$ & CE/ET \\
			2.38 & $1.9\times 10^{-11}$ & $5/3$ & 2 & $0.26$ & $13.5\,\mathrm{km}$ & $1.8\times 10^{3}$ & $-76\%$ & CE/ET \\
			3.48 & $1.3\times 10^{-11}$ & $5/3$ & 2 & $0.26$ & $20\,\mathrm{km}$ & $1.2\times 10^{3}$ & $-76\%$ & CE/ET \\[4pt]
			\multicolumn{9}{@{}l}{\textit{Single-species models, $\gamma_p=5/3$ -- CE/ET band}} \\[2pt]
			0.63 & $5.4\times 10^{-11}$ & $5/3$ & 1 & $0.24$ & $3.9\,\mathrm{km}$ & $6.1\times 10^{3}$ & $-76\%$ & CE/ET \\
			1.09 & $3.1\times 10^{-11}$ & $5/3$ & 1 & $0.24$ & $6.7\,\mathrm{km}$ & $3.5\times 10^{3}$ & $-76\%$ & CE/ET \\
			1.88 & $1.8\times 10^{-11}$ & $5/3$ & 1 & $0.24$ & $11.6\,\mathrm{km}$ & $2.0\times 10^{3}$ & $-76\%$ & CE/ET \\[4pt]
			\multicolumn{9}{@{}l}{\textit{Single-species models, $\gamma_p=2$, $\sigma=5$ -- CE/ET band}} \\[2pt]
			1.00 & $6.6\times 10^{-11}$ & $2$ & 1 & $0.34$ & $4.3\,\mathrm{km}$ & $6.5\times 10^{3}$ & $-82\%$ & CE/ET \\
			2.00 & $3.3\times 10^{-11}$ & $2$ & 1 & $0.34$ & $8.6\,\mathrm{km}$ & $3.2\times 10^{3}$ & $-82\%$ & CE/ET \\
			3.00 & $2.2\times 10^{-11}$ & $2$ & 1 & $0.34$ & $13\,\mathrm{km}$ & $2.2\times 10^{3}$ & $-82\%$ & CE/ET \\[4pt]
			\multicolumn{9}{@{}l}{\textit{Intermediate-mass models, $\gamma_p=4/3$ -- ET band}} \\[2pt]
			10 & $1\times 10^{-12}$ & $4/3$ & 2 & $0.003$ & $5.1\times10^{3}\,\mathrm{km}$ & $0.50$ & $-1\%$ & ET \\
			20 & $1\times 10^{-12}$ & $4/3$ & 2 & $0.012$ & $2.5\times10^{3}\,\mathrm{km}$ & $2.1$ & $-4\%$ & ET \\
			30 & $1\times 10^{-12}$ & $4/3$ & 2 & $0.028$ & $1.6\times10^{3}\,\mathrm{km}$ & $5.1$ & $-9\%$ & ET \\[4pt]
			\multicolumn{9}{@{}l}{\textit{High-compactness models, $\gamma_p=4/3$ -- ET band}} \\[2pt]
			829 & $5.7\times 10^{-14}$ & $4/3$ & 2 & $0.23$ & $5.3\times10^{3}\,\mathrm{km}$ & $4.3$ & $-72\%$ & ET \\
			2\,072 & $2.3\times 10^{-14}$ & $4/3$ & 2 & $0.23$ & $1.3\times10^{4}\,\mathrm{km}$ & $1.7$ & $-72\%$ & ET \\
			4\,144 & $1.1\times 10^{-14}$ & $4/3$ & 2 & $0.23$ & $2.7\times10^{4}\,\mathrm{km}$ & $8.6\times10^{-1}$ & $-72\%$ & ET \\[4pt]
			\multicolumn{9}{@{}l}{\textit{High-compactness models, $\gamma_p=4/3$ -- LISA band}} \\[2pt]
			8\,287 & $5.7\times 10^{-15}$ & $4/3$ & 2 & $0.23$ & $5.3\times10^{4}\,\mathrm{km}$ & $4.3\times10^{-1}$ & $-72\%$ & LISA \\
			$4.1\times10^{4}$ & $1.2\times 10^{-15}$ & $4/3$ & 2 & $0.23$ & $2.6\times10^{5}\,\mathrm{km}$ & $8.7\times10^{-2}$ & $-72\%$ & LISA \\
			$4.1\times10^{5}$ & $1.2\times 10^{-16}$ & $4/3$ & 2 & $0.23$ & $2.6\times10^{6}\,\mathrm{km}$ & $8.7\times10^{-3}$ & $-72\%$ & LISA \\
			$4.1\times10^{6}$ & $1.2\times 10^{-17}$ & $4/3$ & 2 & $0.23$ & $2.6\times10^{7}\,\mathrm{km}$ & $8.7\times10^{-4}$ & $-72\%$ & LISA \\
			$4.1\times10^{7}$ & $1.2\times 10^{-18}$ & $4/3$ & 2 & $0.23$ & $2.6\times10^{8}\,\mathrm{km}$ & $8.7\times10^{-5}$ & $-72\%$ & LISA \\[4pt]
			\bottomrule
		\end{tabular}
	\end{table*}

		Most of the CE/ET-band entries possess radii $R_T\simeq$ $3$--$24\,\mathrm{km}$, the lighter members overlapping the range inferred for neutron stars from NICER X-ray timing \citep{2021ApJ...918L..28M} and the GW170817 tidal-deformability measurement \citep{2018PhRvL.121p1101A}, with contact frequencies $f_{\mathrm{GW}}\sim$ $10^3$--$10^4\,\mathrm{Hz}$ and compactnesses bounded by the genuine maxima, $\kappa\simeq 0.23$--$0.26$ for the soft $\sigma=2$ groups and $\kappa\simeq 0.34$ for the stiff, weak-Bohm group.
	The two-species $\gamma_p=5/3$ sequence is retained as a comparison case to expose the sensitivity of the equilibrium structure to the polytropic index, and yields systematically higher compactness than its $\gamma_p=4/3$ counterpart at the same total mass; the single-species pure quantum-pressure models ($\sigma=0$) reach $\kappa\simeq 0.24$, intermediate between the two.
	The most compact configurations, the stiff, weak-Bohm group ($R_T\sim$ $4$--$13\,\mathrm{km}$, $\kappa\simeq 0.34$) exceed the compactness of canonical $1.4\,\Msun$ neutron stars ($\kappa\approx 0.18$) and cross the photon-sphere value $\kappa=1/3$, acquiring a light ring; they approach, but do not reach, the causal and Buchdahl bounds ($\kappa_{\mathrm{causal}}\simeq 0.35$, $\kappa_{\mathrm{Buchdahl}}=4/9\approx 0.444$ \cite{1959PhRv..116.1027B}).
	From a structural standpoint, configurations with $M\simeq 1$--$2\,\Msun$ and $R_T\simeq 5$--$8\,\mathrm{km}$ would be virtually indistinguishable from neutron stars in gravitational-wave inspiral data, while at the most compact end ($\kappa\sim 0.34$, $R_T/R_S\simeq 1.5$) they cross the photon sphere and could masquerade as low-mass black holes, particularly in the lower mass-gap region ($M\sim 2.5$--$5\,\Msun$) \citep{2020ApJ...896L..44A}.

	Maximum-compactness configurations with $\kappa\approx 0.23$ can also be realised in the ET and LISA bands by admitting sufficiently massive dark-fermion stars: ET-band configurations with $M\sim 10^{3}$--$10^{4}\,\Msun$ ($m_1\sim 10^{-14}\,\eV$, $f_{\mathrm{GW}}\sim 1$--$5\,\mathrm{Hz}$), and LISA-band supermassive configurations with $M\sim 10^{4}$--$10^{7}\,\Msun$ ($m_1\sim 10^{-18}$--$10^{-15}\,\eV$, $f_{\mathrm{GW}}\sim 10^{-4}$--$10^{-1}$$\,\mathrm{Hz}$).
	All retain the $\Delta(M\RT)\approx -72\%$ signature of an intrinsically relativistic equilibrium, mapping the particle mass--stellar mass plane onto contact frequencies spanning some eight decades.
	Figure~\ref{fig:gw_predictions} displays the resulting $f_{\mathrm{GW}}(M)$ curves with the three sensitivity bands overlaid.

	\subsection{Mass--radius relation}

	The mass--radius relation takes the form $M\RT=\mathrm{const}\times\lamB\hbar^2/(Gm_1^2)$, with the constant $M_{\mathrm{dim}}\,\xT$ depending on $\alpha$ (or equivalently on $\kappa$).
	Figure~\ref{fig:mass_radius} displays the equilibrium sequences for five representative particle masses, with the Newtonian ($\alpha=0$) prediction overlaid.
	The GR and Newtonian curves coincide at low compactness ($m_1=10^{-12}\,\eV$, LISA regime) but diverge as the particle mass increases: the departure reaches approximately $10\%$ at $M=1\,\Msun$ for $m_1=5\times10^{-11}\,\eV$ (CE regime) and grows rapidly thereafter.
	The heaviest species plotted ($m_1=10^{-10}\,\eV$) crosses below the neutron-star band already at $M\simeq 0.5\,\Msun$, while the $m_1=5\times10^{-11}\,\eV$ track threads through the $R_T\sim 5$--$10\,\mathrm{km}$ band at $M\simeq 1$--$2\,\Msun$, precisely the locus occupied by neutron stars observed via NICER \citep{2021ApJ...918L..28M}.

	The low-compactness scaling $\RT\propto m_1^{-8/3}$ recovers the expected trend that lighter fermions produce larger, less compact configurations; for $m_1\simeq 3\times10^{-11}\,\eV$ the equilibrium sequence departs from this scaling above $M\simeq 0.3\,\Msun$ and terminates near $\kappa\simeq 0.14$.
	In the stellar-mass regime ($M\sim 1$--$3\,\Msun$), dark-fermion stars with $m_1\sim 3$--$6\times10^{-11}\,\eV$ thread the neutron-star band, reinforcing the mimicker scenario noted above.

	It is natural to ask whether the configurations in Fig.~\ref{fig:mass_radius} are dynamically stable, given that the sequences show no obvious turning point. The relevant test for cold equilibria is the turning-point ($M(\rho_c)$) criterion of \citet{1965gtgc.book.....H} and \citet{1966ApJ...145..505B}, placed on a rigorous footing for rotating stars by \citet{1988ApJ...325..722F}: instability to radial perturbations first appears where $\mathrm{d}M/\mathrm{d}\rho_c=0$. Here $\rho_c=\varepsilon_{\mathrm{tot}}(0)/c^2$ is the central mass--energy density~\eqref{eq:eps_RMF}, which rises monotonically with the compactness parameter $\alpha$, so the criterion applies directly to our $\alpha$-parameterised sequences. Along every sequence the mass rises monotonically, $\mathrm{d}M/\mathrm{d}\rho_c>0$, so all computed models lie on the ascending, stable branch; the absence of a turning point reflects truncation, where the continuation ceases to converge just short of the maximum mass, not marginal stability. The compactness, by contrast, does turn over within the resolved range, at the genuine maxima of Table~\ref{tab:physical}: each sequence reaches its peak compactness, after which the truncation radius reexpands, before the mass maximum is attained. Two features confirm the stable assignment. The equilibria are nodeless ground states, the stable members of such families, as for the lowest branch of fermion and boson stars, whereas nodal (excited) states are generically unstable; and the maximum-mass turning point, beyond which the branch is unstable, lies at central densities above those we resolve, so locating it would require a dedicated radial-pulsation analysis, the relativistic Chandrasekhar equation~\citep{1964ApJ...140..417C} for the lowest $\omega^2$, beyond the scope of this structural survey. We therefore take the sequences to be stable up to, but not including, the unresolved mass maximum. One caveat concerns the most compact members: those crossing the photon sphere ($\kappa>1/3$) are ultracompact horizonless objects, which may in addition suffer the slow trapped-mode (light-ring) instability, a non-radial channel outside the turning-point criterion and likewise left to future work.
	
	\begin{figure}
		\centering
		\includegraphics[width=0.95\columnwidth]{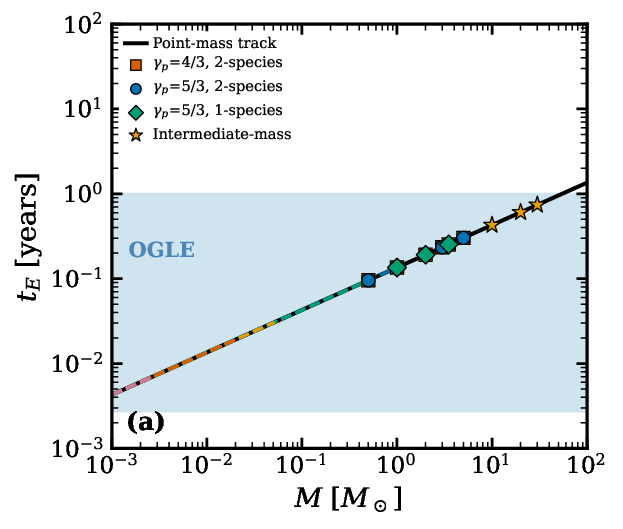}\\[4pt]
		\includegraphics[width=0.95\columnwidth]{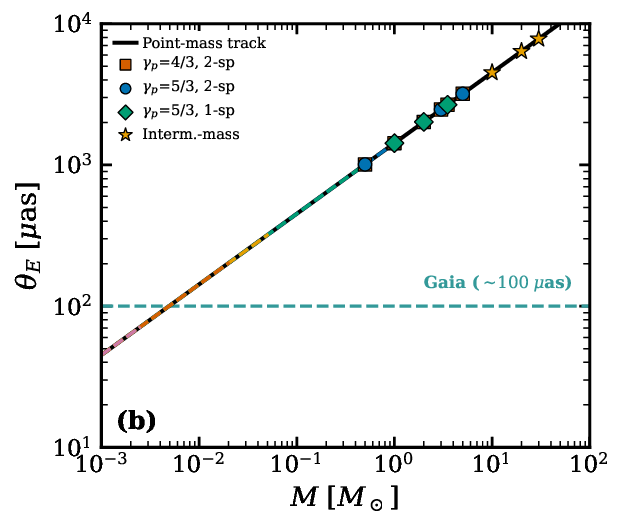}
		\caption{Microlensing predictions for quantum dark fermion stars at lens distance $D_L=4\,\mathrm{kpc}$. (a)~Einstein crossing time $t_E$ as a function of total mass; the shaded band indicates the sensitivity window of the OGLE-IV photometric survey \citep{2015AcA....65....1U}. (b)~Einstein radius $\theta_E$ in microarcseconds; the dashed horizontal line marks the single-epoch astrometric precision of \textit{Gaia} \citep{2022AJ....163..176K}. The solid line shows the universal point-mass track, $\theta_E\propto M^{1/2}$ and $t_E\propto M^{1/2}$. Colored segments correspond to five representative particle masses, $m_1=5\times10^{-13}$, $10^{-12}$, $5\times10^{-12}$, $10^{-11}$, and $5\times10^{-11}\,\eV$, each extending over its equilibrium mass range; dashed segments mark the finite-size regime ($R_T\gtrsim R_E$). Colored markers show the fifteen CE/ET-band configurations and the three intermediate-mass entries ($M=10$--$30\,\Msun$, $m_1=10^{-12}\,\eV$, star symbols) from Table~\ref{tab:physical}.}
		\label{fig:microlensing}
	\end{figure}

	\subsection{Microlensing predictions}
	\label{sec:microlensing}

	Gravitational microlensing surveys afford an independent, electromagnetically based probe of dark compact objects, complementing the gravitational-wave channel \citep{1986ApJ...304....1P}.
	For a point-mass lens of mass $M$ at distance $D_L$, the Einstein radius is $\theta_E=\sqrt{4GM/(c^2 D_L)}$ and the crossing time at transverse velocity $v_\perp\simeq 200\,\mathrm{km\,s}^{-1}$ is $t_E=\theta_E D_L/v_\perp$; the point-mass approximation breaks down when $\RT\gtrsim R_E\equiv\theta_E D_L$, requiring finite-size corrections for the extended configurations at the lighter end of the particle-mass range ($m_1\sim 5\times 10^{-13}\,\eV$).

	Figure~\ref{fig:microlensing} displays $t_E$ and $\theta_E$ as functions of total mass for five representative particle masses, with $D_L=4\,\mathrm{kpc}$ towards the Galactic bulge.
	For $m_1\sim 10^{-12}$--$10^{-11}\,\eV$ and $M\sim 0.5$--$5\,\Msun$, crossing times fall in the range of days to weeks, squarely within the OGLE-IV sensitivity window \citep{2015AcA....65....1U}; heavier particle masses ($m_1\gtrsim 5\times 10^{-11}\,\eV$) yield crossing times of order hours, potentially accessible to high-cadence photometric monitoring.
	The Einstein radii range from approximately $10^3$ to $8\times10^3\,\mu\mathrm{as}$, exceeding the single-epoch astrometric precision of \textit{Gaia} ($\sim\!100\,\mu\mathrm{as}$ for faint bulge sources \cite{2022AJ....163..176K}) by one to two orders of magnitude, so all eighteen Fig.~\ref{fig:microlensing}(b) entries produce individually resolvable astrometric deflections.
	Longer-baseline campaigns such as MOA \citep{2001MNRAS.327..868B} and the Nancy Grace Roman Space Telescope \citep{2019ApJS..241....3P,2022ApJ...933...83S} would extend the reach to configurations outside the parameter range considered here.

	Because higher compactness reduces $\RT$, the point-mass approximation remains valid down to lower particle masses than a low-compactness (Newtonian) estimate would suggest.


	\section{Discussion and conclusions}
	\label{sec:conclusions}

	\subsection{Connection to the dark matter problem}

	If dark matter is composed, even in part, of degenerate fermions in the ultralight regime, then self-gravitating quantum dark fermion stars must exist with properties fixed by a small number of Lagrangian parameters.
	When the Yukawa coupling is comparable to gravity ($\beta_i=1$), the mass--radius relation is entirely determined by the particle mass~$m_1$, yielding a one-parameter predictivity that distinguishes these objects from many exotic compact-object candidates.
	The coupling ratio $\beta_i$ enters as a second independently measurable quantity: a factor-of-ten change in $\beta_i$ shifts the dimensionless mass by more than an order of magnitude (Table~\ref{tab:beta_invariants}), so that a joint measurement of $M$ and $R_T$ constrains both $m_1$ and $\beta_i$ simultaneously, establishing a direct link between gravitational-wave observables and the dark-sector Lagrangian.

	From a structural standpoint, the configurations identified in Table~\ref{tab:physical} are dual mimickers.
	At moderate compactness ($R_T\simeq 5$--$13\,\mathrm{km}$, $M\simeq 0.5$--$3.5\,\Msun$, $\kappa\sim 0.14$--$0.30$) they overlap with neutron stars in mass, radius, and gravitational-wave contact frequency, differing principally in the absence of electromagnetic counterparts: no thermonuclear bursts, no pulsed emission, and no neutrino signal at formation.
	At the most compact end ($\kappa\sim 0.34$, $R_T/R_S\simeq 1.5$, $M\sim 2.5$--$5\,\Msun$), they cross the photon sphere and could masquerade as low-mass black holes in the mass-gap region, where the astrophysical nature of the observed compact remnants remains a matter of debate.
	Distinguishing them from either population would require precise tidal-deformability measurements at high signal-to-noise ratio, or the detection (or demonstrable absence) of an electromagnetic counterpart at merger.

	Three remarks on the scope and reach of the framework are in order. First, the restriction to two species is a physical choice, not a simplification. A single self-coupled species already captures the interplay of degeneracy pressure, the Bohm quantum term, and the Yukawa fifth force; two are the minimal setting in which the components segregate, the heavier forming a compact core and the lighter an extended envelope (Fig.~\ref{fig:two_species}), as expected in asymmetric or multicomponent dark matter with a heavy ``dark baryon'' and a lighter partner. The two-component star is thus the simplest member of a broader family. Second, that family extends naturally to $N$ species coupled through one or more mediators: each additional Dirac field brings its own modified Tolman--Oppenheimer--Volkoff equation~\eqref{eq:TOV_bohm} and Walecka source in the Klein--Gordon equation~\eqref{eq:KG_nuclear}, and $N-1$ mediators promote the single coupling ratio $\beta_i$ to a matrix $\beta_{ij}$, leaving the structure of the boundary-value problem unchanged. Only the numerical cost grows, as for the multistate and multifield generalisations of boson stars \citep{2010PhRvD..81d4031B,2023LRR....26....1L}; these extensions, and the richer mass--radius phenomenology they permit, are natural directions for future work. Third, the forward solutions already delimit the compactness the framework can reach: the genuine maxima span $\kappa\simeq0.23$--$0.34$, the stiffest, weak-Bohm configurations ($\gamma_p=2$, $\sigma=5$) reaching the upper end and crossing the photon-sphere value $\kappa=1/3$. This ceiling is robust on two counts: it is set by the causal stiffness limit of the equation of state ($\gamma_p\to2$, $c_s\to c$), and adding the second species does not raise it, the two-fluid maxima recovering the single-species value as the lighter component is diluted and falling below it otherwise. Within the isotropy assumed throughout, these are the maximally compact, light-ring-bearing members of the family; surpassing them towards the Buchdahl bound $\kappa=4/9$ would require anisotropic support, beyond the present scope.

	\subsection{Observational prospects}

	Gravitational-wave inspiral signals from binary mergers constitute the most direct probe.
	Tidal deformability measurements during the late inspiral are particularly powerful: because quantum dark fermion stars are supported by degeneracy pressure together with the globally relevant Bohm contribution rather than by nuclear forces, their dimensionless tidal deformability $\Lambda$ differs measurably from that of neutron stars at comparable compactness \citep{2010PhRvD..81l3521D,2026PhRvD.113d3049B}.
	In the BH-mimicker regime ($\kappa\gtrsim 0.3$, $R_T\lesssim 10\,\mathrm{km}$), $\Lambda$ is much smaller than the neutron-star value but remains nonzero, distinguishing these configurations from genuine black holes ($\Lambda=0$).
	Tidal-deformability measurements thus discriminate against both populations simultaneously.

	Complementary constraints arise from gravitational microlensing, astrometric perturbations observed by \textit{Gaia}, pulsar-timing residuals, and continuous gravitational-wave searches \citep{2015AcA....65....1U,2001MNRAS.327..868B,2022ApJ...933...83S,2019PhRvD.100b3003D,2025PhRvD.111d3009B}.
	The absence of an electromagnetic counterpart in a well-localised lensing event would constitute circumstantial evidence for a dark compact lens.
	Taken together with the gravitational-wave channel, these probes define a multimessenger strategy for constraining the dark-sector parameter space.

	\subsection{Summary}

	We have developed a general-relativistic framework for two-component quantum dark fermion stars, configurations in which equilibrium is governed jointly by gravity, a Yukawa-mediated dark fifth force, and a globally relevant Bohm quantum-pressure correction.
	The framework retains the full covariant treatment of the nonlinear Klein--Gordon equation in the Schwarzschild interior and admits closure relations valid at arbitrary compactness, from the ultralight baseline through to the nuclear-density regime where relativistic scalar densities and self-consistent effective fermion masses are required.

	Two Lagrangian parameters, the particle mass and the coupling ratio between the Yukawa channel and gravity, together fix the equilibrium structure; a joint measurement of mass and radius then constrains the dark-sector microphysics directly from gravitational-wave observables.
	The framework is markedly sensitive to the second parameter: across the surveyed range $\beta_i=0.5$ to $10$, the dimensionless equilibrium mass varies by nearly two orders of magnitude.

	The principal physical finding is that ultralight degenerate fermions yield equilibrium configurations whose radii and compactnesses span the neutron-star regime and extend into the strongly relativistic limit, where the most compact members approach the Schwarzschild radius closely enough to mimic low-mass black holes.
	The same family therefore behaves as dual mimickers: at moderate compactness it overlaps with neutron stars in mass, radius, and inspiral frequency; at the most compact end it could masquerade as low-mass black holes in the mass-gap region ($M\sim 2.5$--$5\,\Msun$).
	Tidal-deformability measurements discriminate against both populations simultaneously, being measurably smaller than the neutron-star value yet remaining nonzero, unlike that of a genuine black hole.

	The full set of equilibria populates the sensitivity bands of LISA, the Einstein Telescope, and Cosmic Explorer, with contact gravitational-wave frequencies spanning some eight decades, from $\sim 10^{-4}\,\mathrm{Hz}$ for supermassive configurations ($M\sim 10^{7}\,\Msun$) to $\sim 10^{4}\,\mathrm{Hz}$ for stellar-mass entries ($M\sim 0.5\,\Msun$), and the most compact members produce astrometric microlensing signatures individually resolvable by \textit{Gaia}.
	These results establish two-component quantum dark fermion stars as a reproducible and falsifiable class of compact objects, whose equilibrium structure encodes the dark-sector Lagrangian in macroscopic, multimessenger observables.

	\begin{acknowledgments}
		I.~L. thanks the anonymous referee for a careful and constructive reading of the manuscript, whose comments have improved the quality and clarity of the presentation. I.~L. thanks the Fundação para a Ciência e Tecnologia (FCT), Portugal, for the financial support to the Center for Astrophysics and Gravitation (CENTRA/IST/ULisboa) through grant No.\ UID/PRR/00099/2025 (\doi{10.54499/UID/PRR/00099/2025}) and grant No.\ UID/00099/2025 (\doi{10.54499/UID/00099/2025}).
	\end{acknowledgments}

	\bibliography{ArtILopes}

	\appendix

	\section{Summary of notation and dimensionless variables}
	\label{app:notation}

	The notation employed throughout is summarised here for convenience: Table~\ref{tab:notation_phys} gathers the physical (dimensionful) quantities, each with its defining equation, while Table~\ref{tab:notation_dimless} collects the rescaled variables and dimensionless parameters that govern the boundary-value problem.

	\begin{table}[htbp]
		\caption{Physical (dimensionful) quantities. The defining equation, where applicable, is given in the final column.}
		\label{tab:notation_phys}
		\begin{ruledtabular}
		\begin{tabular}{cll}
			Symbol & Meaning & Eq. \\
			\colrule
			$g_{\mu\nu}$ & spacetime metric, signature $(-,+,+,+)$ & \eqref{eq:metric} \\
			$R_{\mu\nu},\,R$ & Ricci tensor and Ricci scalar & \eqref{eq:Einstein} \\
			$e^{\ a}_{\mu},\,\Omega_\mu$ & vierbein and spin connection & \eqref{eq:Lagrangian} \\
			$m_1,\,m_2$ & rest masses of species $1$ and $2$ & \eqref{eq:Tmunu_decomp} \\
			$\mphi$ & mediator (scalar) mass & \eqref{eq:Vphi} \\
			$\gphi,\,\gchi$ & Yukawa couplings of species $1,2$ & \eqref{eq:Lagrangian} \\
			$\phi$ & scalar mediator field & \eqref{eq:Lagrangian} \\
			$m_i^*(r)$ & effective (Walecka) mass & \eqref{eq:mstar_def} \\
			$n_i,\,n_{s,i}$ & number and scalar densities & \eqref{eq:ns_walecka} \\
			$\varepsilon_{\mathrm{tot}},\,P_{\mathrm{tot}}$ & total energy density and pressure & \eqref{eq:eps_RMF} \\
			$\rho_c$ & central mass--energy density $\varepsilon_{\mathrm{tot}}(0)/c^2$ & \eqref{eq:eps_RMF} \\
			$Q_i^{\mathrm{GR}}$ & curved-spacetime Bohm potential & \eqref{eq:bohm_GR} \\
			$\lamB$ & Kirzhnits gradient coefficient ($=1/9$) & \eqref{eq:bohm_GR} \\
			$\beta_i$ & Yukawa-to-gravity coupling ratio & \eqref{eq:beta_def} \\
			$m(r),\,M$ & enclosed and total gravitational mass & \eqref{eq:mass_eq} \\
			$\RT,\,\kappa$ & total radius and compactness $GM/\RT c^2$ & \eqref{eq:kappa_physical} \\
		\end{tabular}
		\end{ruledtabular}
	\end{table}

	\begin{table}[htbp]
		\caption{Dimensionless variables and parameters of the boundary-value problem of Sections~\ref{sec:dimensionless}--\ref{sec:numerics}.}
		\label{tab:notation_dimless}
		\begin{ruledtabular}
		\begin{tabular}{cll}
			Symbol & Meaning & Eq. \\
			\colrule
			$x$ & dimensionless radius $r/r_0$ & \eqref{eq:eta_s_def} \\
			$\eta_{s,i}$ & scalar-density amplitude $\sqrt{n_{s,i}/n_0}$ & \eqref{eq:eta_s_def} \\
			$\eta_i$ & number-density amplitude $\sqrt{n_i/n_0}$ & \eqref{eq:eta_s_def} \\
			$\varphi$ & normalized mediator/gravitational potential & \eqref{eq:scales} \\
			$\Phi_0,\,\psi_0^2$ & energy and density scales & \eqref{eq:scales} \\
			$\sigma_i$ & dimensionless polytropic pressure parameter & \eqref{eq:sigma_def} \\
			$\xi$ & dimensionless mediator mass $\kappaphi r_0$ & \eqref{eq:sigma_def} \\
			$\alpha$ & compactness parameter $\Phi_0/c^2$ & \eqref{eq:alpha_def} \\
			$q$ & species mass ratio $m_1/m_2\geq 1$ & \eqref{eq:gr_eta2} \\
			$f$ & central density fraction of species $2$ & \eqref{eq:BC_GR} \\
			$\chi_0$ & central relativity parameter $p_F/m c$ & \eqref{eq:etas_eta_inversion} \\
			$\tilde\kappa_3,\,\tilde\lambda_\phi$ & dimensionless cubic, quartic self-couplings & \eqref{eq:dimless_selfcouplings} \\
			$A(x)$ & dimensionless metric function & \eqref{eq:cov_correction} \\
			$\mathcal{M}_c(x)$ & dimensionless enclosed mass & \eqref{eq:Mdim} \\
			$M_{\mathrm{dim}},\,\xT$ & dimensionless total mass and radius & \eqref{eq:MR_physical} \\
			$\mathcal{E}_i,\,v_c$ & density and potential eigenvalues & \eqref{eq:BC_GR} \\
		\end{tabular}
		\end{ruledtabular}
	\end{table}

	\section{Derivation of the Bohm force from the stress-energy tensor}
	\label{app:bohm_derivation}

	This appendix derives the Bohm quantum-pressure term $-n_i\,\mathrm{d}Q_i^{\mathrm{GR}}/\mathrm{d}r$ in the modified TOV equation, starting from covariant conservation of the stress-energy tensor.

	\subsection{Decomposition of the fermion stress-energy tensor}

	In the orbital-free density-functional framework, the stress-energy tensor for each fermion species decomposes as
	\begin{equation}
		T_{\mu\nu}^{(i)} = \underbrace{(\varepsilon_i+P_i)\,\frac{u_\mu u_\nu}{c^2}+P_i\,g_{\mu\nu}}_{T_{\mu\nu}^{(\mathrm{pf},i)}} + \;\Pi_{\mu\nu}^{(Q,i)}\,,
		\label{eq:Tmunu_decomp_species}
	\end{equation}
	where $T_{\mu\nu}^{(\mathrm{pf},i)}$ is the standard perfect-fluid contribution and $\Pi_{\mu\nu}^{(Q,i)}$ is the quantum stress tensor from the Weizs\"acker--Kirzhnits gradient correction \citep{1935ZPhy...96..431W,2018PhPl...25c2115M,2025PhyU...68..691P}.

	\subsection{The quantum stress tensor}

	The Weizsäcker gradient correction to the Thomas--Fermi kinetic-energy density is
	\begin{equation}
		t_W^{(i)} = \frac{\lamB\hbar^2}{8m_i}\,g^{\alpha\beta}\,\frac{\nabla_\alpha n_i\,\nabla_\beta n_i}{n_i}\,,
		\label{eq:tW}
	\end{equation}
	where $\lamB=1/9$ is the Kirzhnits coefficient.
	Variation of the corresponding action with respect to the metric yields the quantum stress tensor
	\begin{equation}
		\Pi_{\mu\nu}^{(Q,i)} = \frac{\lamB\hbar^2}{4m_i}\!\left(\frac{\nabla_\mu n_i\,\nabla_\nu n_i}{n_i}-\nabla_\mu\nabla_\nu n_i\right) + g_{\mu\nu}\,\mathcal{U}_Q^{(i)}\,,
		\label{eq:Pi_Q}
	\end{equation}
	where
	\begin{equation}
		\mathcal{U}_Q^{(i)} = \frac{\lamB\hbar^2}{4m_i}\!\left(\nabla^\alpha\nabla_\alpha n_i - \frac{1}{2}\,\frac{g^{\alpha\beta}\nabla_\alpha n_i\,\nabla_\beta n_i}{n_i}\right)
		\label{eq:UQ}
	\end{equation}
	is the covariant generalisation of the flat-space quantum stress tensor.

	\subsection{Covariant divergence and the Bohm potential}

	We now compute the covariant divergence of the quantum stress tensor $\Pi_{\mu\nu}^{(Q,i)}$ defined in Eq.~\eqref{eq:Pi_Q}, and identify its compact form in terms of the curved-spacetime Bohm potential.
	The divergence of the first (trace-free) part is
	\begin{equation}
		\nabla^\mu\!\left(\frac{\nabla_\mu n_i\,\nabla_\nu n_i}{n_i}-\nabla_\mu\nabla_\nu n_i\right) = \frac{\nabla_\nu n_i\,\nabla^2_{\mathrm{cov}} n_i}{n_i}-\frac{(\nabla^\alpha n_i\,\nabla_\alpha n_i)\,\nabla_\nu n_i}{n_i^2}\,.
	\end{equation}
	The divergence of the $g_{\mu\nu}\mathcal{U}_Q^{(i)}$ part contributes
	\begin{equation}
		\nabla_\nu\mathcal{U}_Q^{(i)} = \frac{\lamB\hbar^2}{4m_i}\,\nabla_\nu\!\left(\nabla^2_{\mathrm{cov}} n_i - \frac{|\nabla n_i|^2}{2n_i}\right),
	\end{equation}
	where $|\nabla n_i|^2\equiv g^{\alpha\beta}\nabla_\alpha n_i\,\nabla_\beta n_i$.
	The two contributions may be recombined using the chain-rule identity
	\begin{equation}
		\frac{\nabla^2_{\mathrm{cov}}\!\sqrt{n_i}}{\sqrt{n_i}} = \frac{\nabla^2_{\mathrm{cov}} n_i}{2n_i} - \frac{|\nabla n_i|^2}{4n_i^2}\,.
		\label{eq:chain_rule}
	\end{equation}
	Adding the trace-free and trace parts and applying Eq.~\eqref{eq:chain_rule} gives the compact result
	\begin{equation}
		\nabla^\mu\Pi_{\mu\nu}^{(Q,i)} = n_i\,\nabla_\nu Q_i^{\mathrm{GR}}\,,
		\label{eq:div_Pi}
	\end{equation}
	where
	\begin{equation}
		Q_i^{\mathrm{GR}} \equiv -\frac{\lamB\hbar^2}{2m_i}\,\frac{\nabla^2_{\mathrm{cov}}\!\sqrt{n_i}}{\sqrt{n_i}}
		\label{eq:Q_GR_def}
	\end{equation}
	is the curved-spacetime Bohm potential, emerging naturally from the algebra rather than postulated.
	In the static, spherically symmetric case considered throughout this work, $Q_i^{\mathrm{GR}}$ is a scalar function of $r$ alone, so $\nabla_\nu Q_i^{\mathrm{GR}}$ has a single nonzero component and Eq.~\eqref{eq:div_Pi} reduces to
	\begin{equation}
		\nabla^\mu\Pi_{\mu r}^{(Q,i)} = n_i\,\frac{\mathrm{d}Q_i^{\mathrm{GR}}}{\mathrm{d}r}\,,
		\label{eq:div_Pi_radial}
	\end{equation}
	which is the form invoked in the recovery of the modified TOV equation below.

	\subsection{Recovery of the modified TOV equation}

	Covariant conservation of each species' stress-energy tensor, $\nabla^\mu T_{\mu\nu}^{(i)}=0$, applied to the decomposition~\eqref{eq:Tmunu_decomp_species}, yields
	\begin{equation}
		\underbrace{\nabla^\mu T_{\mu\nu}^{(\mathrm{pf},i)}}_{\displaystyle\frac{\mathrm{d}P_i}{\mathrm{d}r}+(\varepsilon_i+P_i)\frac{\mathrm{d}\nu}{\mathrm{d}r}} + \;\underbrace{\nabla^\mu\Pi_{\mu\nu}^{(Q,i)}}_{\displaystyle n_i\frac{\mathrm{d}Q_i^{\mathrm{GR}}}{\mathrm{d}r}} = 0\,,
	\end{equation}
	where the first term is the standard perfect-fluid divergence in the static metric, and the second term follows from Eq.~\eqref{eq:div_Pi}.
	Rearranging gives the modified TOV equation,
	\begin{equation}
		\frac{\mathrm{d}P_i}{\mathrm{d}r} = -(\varepsilon_i+P_i)\,\frac{\mathrm{d}\nu}{\mathrm{d}r} - n_i\,\frac{\mathrm{d}Q_i^{\mathrm{GR}}}{\mathrm{d}r}\,,
	\end{equation}
	demonstrating that the Bohm force term originates from the quantum correction to the perfect-fluid stress-energy tensor.
	
\smallskip

	\section{Recovery of the number density from the scalar density}
	\label{app:walecka_inversion}

	The boundary-value problem of Section~\ref{sec:dimensionless} carries the scalar-density amplitude $\eta_{s,i}$ as a state variable, whereas the polytropic pressure terms are expressed through the number-density amplitude $\eta_i$; the two are connected by the Walecka relation, which we now invert to recover $\eta_i$ from $\eta_{s,i}$.
	To make this relation explicit, we substitute $t=p/m_i^*$ in Eq.~\eqref{eq:ns_walecka} and define $t_{F,i}=p_{F,i}/m_i^*$, giving
	\begin{equation}
		n_{s,i} = \frac{{m_i^*}^3}{\pi^2}\,I(t_{F,i})\,,\qquad
		I(t_F)=\tfrac{1}{2}\bigl[t_F\sqrt{t_F^2+1}-\operatorname{arcsinh}(t_F)\bigr]\,.
		\label{eq:walecka_I}
	\end{equation}
	Dividing by $n_i=p_{F,i}^3/(3\pi^2)={m_i^*}^3\,t_{F,i}^3/(3\pi^2)$ yields the ratio
	\begin{equation}
		R(t_F)\equiv\frac{n_{s,i}}{n_i}=\frac{3\,I(t_F)}{t_F^3}\,,
		\label{eq:walecka_R}
	\end{equation}
	whence
	\begin{equation}
		\eta_{s,i}^2 = \eta_i^2\,R(t_{F,i})\,,\qquad
		t_{F,i}=\frac{\chi_0\,\eta_i^{2/3}}{m_i^*/m_i}\,,
		\label{eq:etas_eta_inversion}
	\end{equation}
	with $\chi_0=(3\pi^2 n_0)^{1/3}\hbar/(m_i c)$ the central relativity parameter.
	Here $n_0$ is the reference central density scale fixed by the normalization $\eta_{s,1}(0)=1$; the true central number density of species~$i$ is $n_{i,c}=n_0\,\eta_i^2(0)$, which differs from $n_0$ whenever $\chi_0>0$ because the Walecka inversion yields $\eta_i(0)\neq\eta_{s,i}(0)$.
	Equivalently, $\chi_0$ may be expressed in terms of $n_{i,c}$ as $\chi_0 = (3\pi^2 n_{i,c})^{1/3}\hbar/(m_i c\,\eta_i^{2/3}(0))$, making explicit that it encodes the ratio of the central Fermi momentum to the rest energy.
	The solver inverts Eq.~\eqref{eq:etas_eta_inversion} by Newton iteration at each grid point: given $\eta_{s,i}$ from the BVP state vector, it finds the unique $\eta_i$ satisfying $\eta_{s,i}^2 = \eta_i^2\,R\!\bigl(\chi_0\,\eta_i^{2/3}/(m_i^*/m_i)\bigr)$.
	When $\chi_0=0$, $R\to 1$ and the inversion reduces to the identity $\eta_i=\eta_{s,i}$.

\end{document}